\documentclass{llncs}

\usepackage{graphicx}
\usepackage{color,slashbox,colortbl}
\usepackage{amssymb,amsmath, stmaryrd}
\usepackage{isabelle,isabellesym}
\usepackage{subfig}
\usepackage{caption}

\usepackage{pgf,tikz}
\usetikzlibrary{arrows,chains,matrix,positioning,scopes,graphs}

\title{A Proof of the Compositions of \\Time Interval Relations}
 
\author{
Fadoua Ghourabi\inst{1}
\and
    Kazuko Takahashi\inst{2}}

\institute{
  Ochanomizu University,
  Tokyo, Japan\\
  \email{ghourabi.fadoua@ocha.ac.jp}
\and
   Kwansei Gakuin University,
   Sanda, Japan.\\
   \email{ktaka@kwansei.ac.jp}
 }

\authorrunning{Ghourabi and Takahashi}

\begin{document}

\maketitle

\begin{abstract}
We prove  the 169 compositions of time interval relations. The proof is first-order and inferred from an axiomatic system on time intervals. We show a general proof template that can  alleviate the manual proof with Isar. 
 \end{abstract}

\section{Introduction}
Allen's interval calculus is a qualitative knowledge representation formalism in the first order logic~\cite{Allen:1983,Allen:1985}. It is motivated by the qualitative verbalization of events in day to day conversation where we are not that much concerned about  dates and durations, i.e.~numerical data. It inspired rethinking the way objects in the space are represented which yield to several qualitative spatial representations~\cite{Randell:1992a,Frank:1991}. 

A qualitative representation is generally based on a relation algebra. Allen introduces 13 binary relations that define all possible arrangements that can exist between two events. Two events can be \textit{before}, \textit{equal}, \textit{starting}, \textit{finishing}, \textit{overlapping}, \textit{meeting} or \textit{during} each other.  
The compositions of Allen's relations are pertinent to the reasoning about knowledge of time. In particular, a consistency problem of relation constraints is commonly solved with a guideline from these compositions~\cite{Ladkin:1992}. 
All the 169 compositions are first given in~\cite{Allen:1983}. 


Wolter and Dylla \cite{Wolter:2012,Dylla:2013} designed an algebraic framework that define the qualitative relations in an algebraic domain. The framework is independent of the calculus and 
 covers several qualitative spatial and temporal systems including Allen's interval calculus. 
 The relation constraints are solved using algebraic automated proving methods such as Gr\"{o}bner basis (GB)~\cite{Buchberger:1985} and cylindrical algebraic decomposition (CAD)~\cite{Collins:1996}. 
 The algebraic framework of qualitative calculus is implemented in SparQ tool~\cite{Wallgrun:2007}. The user can perform all sort of manoeuvres, e.g. checking the consistency of relation constraints, computing an algebraic closure, generating a quantitative scenario from qualitative data and vice versa, etc. 

The computation of GB and CAD is a challenging task affected by the number of variables and the degree of algebraic constraints. Checking compositions  of Allen's interval calculus is, however, not exciting.  Time interval relations are interpreted in the 1D. The algebraic constraints of compositions are inequalities of degree 1 with at most 6 variables. 
Algebraic methods in a computer algebra system are powerful enough to check these algebraic constraints in milliseconds.

The logically inferred proof of validity of the compositions  that is independent of the interpretation domain has yet to be done. We proved  the  compositions  of Allen's relations with Isar, and in this paper, we explain how we proceeded to that end. 
When proving methods or formalism in the qualitative knowledge,  the design of a proof strategy is equally important than the result of the proof. The issue that arises is that the number of cases is huge and proving properties about them is cumbersome in an  interactive proof style. In \cite{Fadoua:2015}, we handled this situation by grouping cases  into equivalent classes and then showing that it is enough to prove the properties for a representative case of each class. In this paper, the ordering of the relations in a lattice gives direction on the general steps of the proofs. We  design a kind of ``template" to structure the proofs. We can either use it as it is or extend it depending on the composition to prove.  

The rest of the paper is organized as follows. 
In Sect.~\ref{sect:arel}, we introduce the basic time interval relations. In Sect. \ref{sect:axioms}, we present the formalization of the axiomatic system. In Sect.~\ref{sect:arel} and \ref{sect:jepd} we present the formalization of time interval relations and show their properties. We introduce the composition table in Sect.~\ref{sect:comp}, then in Sect.~\ref{sect:proof} we explain its proof with Isar. In Sect.~\ref{sect:conc}, we conclude with remarks on future directions of research.
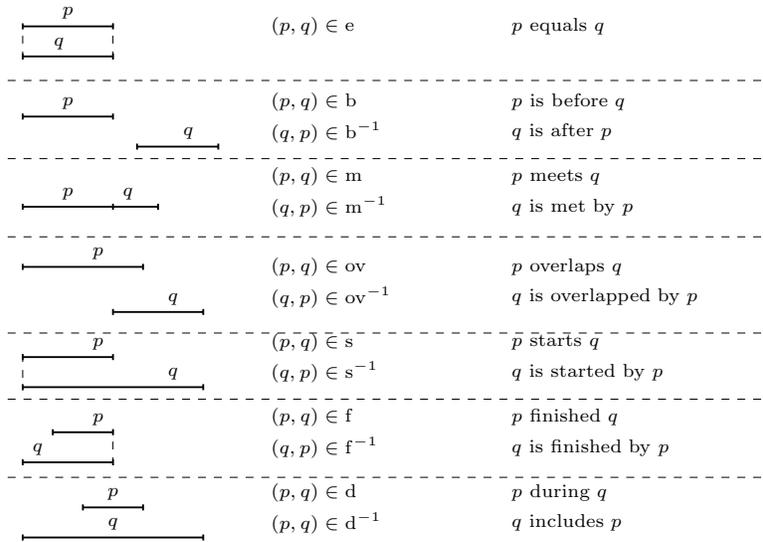
\begin{figure}[h]
\centering
\begin{tikzpicture}[x=.5cm,y=.5cm, scale =0.8]
\clip(-0.5,-16) rectangle (25,4);
\scriptsize
\draw [line width = 0.7pt] (0,2) -- (3,2);
\draw[line width = 0.7pt] (0,1.9) -- (0,2.1);
\draw[line width = 0.7pt] (3,1.9) -- (3,2.1);
\node [above] at (1.5,2) { $p$};
\draw [line width = 0.7pt] (0,1) -- (3,1);
\draw [line width = 0.7pt] (0,0.9) -- (0,1.1);
\draw [line width = 0.7pt] (3,0.9) -- (3,1.1);
\node [above ] at (1.2, 1) { $q$};
\draw [line width = 0.2pt, dashed] (0,0.9) -- (0,1.9);
\draw [line width = 0.2pt, dashed] (3,0.9) -- (3,2.1);

\node [right] at (8,2) {$(p,q) \in \mathrm{e}$};
\node [right] at (16,2) {$p$ {equals} $q$};

\draw[line width= 0.2pt, dashed] (-0.5,0.2) -- (25,0.2);
\draw [line width = 0.7pt] (0,-1) -- (3,-1);
\draw[line width = 0.7pt] (0,-1.1) -- (0,-0.9);
\draw[line width = 0.7pt] (3,-1.1) -- (3,-0.9);
\node [above] at (1.5,-1) { $p$};
\draw [line width = 0.7pt] (3.8,-2) -- (6.5,-2);
\draw [line width = 0.7pt] (3.8,-2.1) -- (3.8,-1.9);
\draw [line width = 0.7pt] (6.5,-2.1) -- (6.5,-1.9);
\node [above] at (5.5,-2) { $q$};
\node [right] at (8,-0.5) {$(p,q) \in \mathrm{b}$};
\node [right] at (16,-0.5) {$p$ {is before} $q$};
\node [right] at (8,-1.5) {$(q,p) \in \mathrm{b}^{-1}$};
\node [right] at (16,-1.5) {$q$ {is after} $p$};

\draw[line width= 0.2pt, dashed] (-0.5,-2.4) -- (25,-2.4);

\draw [line width = 0.7pt] (0,-4) -- (3,-4);
\draw [line width = 0.7pt] (0,-4.1) -- (0,-3.9);
\draw [line width = 0.7pt] (3,-4.1) -- (3,-3.9);
\node [above] at (1.5,-4) { $p$};
\draw [line width = 0.7pt] (3,-4) -- (4.5,-4);
\draw [line width = 0.7pt] (4.5,-4.1) -- (4.5,-3.9);
\node [above] at (3.5,-4) { $q$};
\node [right] at (8,-3) {$(p,q) \in \mathrm{m}$};

\node [right] at (8,-4) {$(q,p) \in \mathrm{m}^{-1}$};
\node [right] at (16,-3) {$p$ {meets} $q$};
\node [right] at (16,-4) {$q$ {is met by} $p$};

\draw[line width= 0.2pt, dashed] (-0.5,-5) -- (25,-5);

\draw [line width = 0.7pt] (0,-6) -- (4,-6);
\draw [line width = 0.7pt] (0,-6.1) -- (0,-5.9);
\draw [line width = 0.7pt] (4,-6.1) -- (4,-5.9);
\node [above] at (2.5,-6) { $p$};
\draw [line width = 0.7pt] (3,-7.5) -- (6,-7.5);
\draw [line width = 0.7pt] (3,-7.6) -- (3,-7.4);
\draw [line width = 0.7pt] (6,-7.6) -- (6,-7.4);
\node [above] at (5,-7.5) { $q$};
\node [right] at (8,-6) {$(p,q) \in \mathrm{ov}$};

\node [right] at (8,-7) {$(q,p) \in \mathrm{ov}^{-1}$};
\node [right] at (16,-6) {$p$ overlaps $q$};
\node [right] at (16,-7) {$q$ {is overlapped by} $p$};

\draw[line width = 0.2pt, dashed] (-0.5, -8.2) -- (25,-8.2);

\draw [line width = 0.7pt] (0,-9) -- (3,-9);
\draw [line width = 0.7pt] (0,-9.1) -- (0,-8.9);
\draw [line width = 0.7pt] (3,-9.1) -- (3,-8.9);
\node [above] at (2.5,-9) { $p$};
\draw [line width = 0.7pt] (0,-10) -- (6,-10);
\draw [line width = 0.7pt] (0,-10.1) -- (0,-9.9);
\draw [line width = 0.7pt] (6,-10.1) -- (6,-9.9);
\node [above] at (5,-10) { $q$};

\draw[line width = 0.2pt, dashed] (0,-8.9) -- (0,-10.1);
\node [right] at (8,-8.5) {$(p,q) \in \mathrm{s}$};

\node [right] at (8,-9.5) {$(q,p) \in \mathrm{s}^{-1}$};
\node [right] at (16,-8.5) {$p$ starts $q$};
\node [right] at (16,-9.5) {$q$ {is started by} $p$};

\draw[line width = 0.2pt, dashed] (-0.5, -10.4) -- (25,-10.4);


\draw [line width = 0.7pt] (1,-11.5) -- (3,-11.5);
\draw [line width = 0.7pt] (1,-11.6) -- (1,-11.4);
\draw [line width = 0.7pt] (3,-11.6) -- (3,-11.4);
\node [above] at (2.5,-11.5) { $p$};
\draw [line width = 0.7pt] (0,-12.5) -- (3,-12.5);
\draw [line width = 0.7pt] (0,-12.6) -- (0,-12.4);
\draw [line width = 0.7pt] (3,-12.6) -- (3,-12.4);
\node [above] at (0.5,-12.5) { $q$};

\draw[line width = 0.2pt, dashed] (3,-12.6) -- (3,-11.4);
\node [right] at (8,-11) {$(p,q) \in \mathrm{f}$};

\node [right] at (8,-12) {$(q,p) \in \mathrm{f}^{-1}$};
\node [right] at (16,-11) {$p$ finished $q$};
\node [right] at (16,-12) {$q$ {is finished by} $p$};

\draw[line width = 0.2pt, dashed] (-0.5, -13) -- (25,-13);


\draw [line width = 0.7pt] (2,-14) -- (4,-14);
\draw [line width = 0.7pt] (2,-14.1) -- (2,-13.9);
\draw [line width = 0.7pt] (4,-14.1) -- (4,-13.9);
\node [above] at (3,-14) { $p$};
\draw [line width = 0.7pt] (0,-15) -- (6,-15);
\draw [line width = 0.7pt] (0,-15.1) -- (0,-14.9);
\draw [line width = 0.7pt] (6,-15.1) -- (6,-14.9);
\node [above] at (3,-15) { $q$};
\node [right] at (8,-13.5) {$(p,q) \in \mathrm{d}$};

\node [right] at (8,-14.5) {$(p,q) \in \mathrm{d}^{-1}$};
\node [right] at (16,-13.5) {$p$ during $q$};
\node [right] at (16,-14.5) {$q$ {includes} $p$};
\end{tikzpicture}
\caption{The 13 time interval relations}
\label{fig:arels}
\end{figure}

\section{Basic Relations}
\label{sect:brel}

An event is continuous in a finite period of time. An event is qualitatively represented as a time interval, and  the basic objects that we consider in our formalization are  intervals. 
For simplicity (and like most papers on Allen's calculus), we illustrate our explanation with a spatial representation of intervals as parallel line segments. Moreover,  the notions of ``starting point" and ``ending point" of intervals- although not formally defined- support our explanation. 

There are  13 possible arrangements of time intervals. They can be described with relations {before}, {equal}, {starting}, {finishing}, {overlapping}, {meeting} or {during}. 
The relations together with their respective inverses are 13 and they are depicted in Fig.~\ref{fig:arels}. Hereafter, relations before, meets, overlaps, starts, finishes, during and equal are abbreviated to \texttt{b}, \texttt{m}, \texttt{ov}, \texttt{s}, \texttt{f}, \texttt{d} and \texttt{e}, respectively. 

When the relation between two events is completely known, then it is represented with one of the 13 time interval relations that we call \textit{basic} relations. For example, the statement  ``Tarski (T) lived after Euclid (E)."  unambiguously expresses that Euclid's life preceded Tarski's life, which can be represented (T, E) $\in$ \texttt{b}$^{-1}$ (or (E, T) $\in$ \texttt{b}). 

Sometimes, the knowledge is not precise. The statement ``Hilbert (H) was born before Tarski (T)." lacks information about who died first and whether one was born after the death of the other, etc.  Nevertheless, we still can  represent these various situations as follows (H, T) $\in$ \texttt{b} $\cup$ \texttt{ov} $\cup$ \texttt{m} $\cup$ \texttt{d}$^{-1}$ $\cup$ \texttt{f}$^{-1}$, i.e.~the life of Hilbert is either before or overlaps or meets or includes or finishes by the life of Tarski. We can derive new qualitative knowledge by composing relations. The relation (H, E) is the composition of relations  \texttt{b} $\cup$ \texttt{ov} $\cup$ \texttt{m} $\cup$ \texttt{d}$^{-1}$ $\cup$ \texttt{f}$^{-1}$ and \texttt{b}$^{-1}$.

For simplicity, $(r_1, \ldots, r_n)$ denotes the union of basic relations $r_1, \ldots, r_n$. Thus, $x \in (r_1, \ldots, r_n)$ is equivalent to $x \in r_1 \lor \ldots \lor  x \in r_n$. Let $\alpha_1 = (r_1, \ldots, r_i)$ and $\alpha_2 = (r_1', \ldots, r_j')$, we denote by $\alpha_1 + \alpha_2$ the union $(r_1, \ldots, r_i, r_1', \ldots, r_j')$. 

\section{Axioms}
\label{sect:axioms}

\begin{figure}[h]
\centering
\begin{tikzpicture}[line cap=round,line join=round,>=triangle 45,x=1.0cm,y=1.0cm, scale = 0.5]
\clip(-1,0) rectangle (6,2);

\node [above] at (1.5,1) { $p$};
\draw [black, line width = 1.2pt ] (0,1) -- (3,1);
\draw [black , line width = 1.2pt ] (0,1.2) -- (0,0.8);
\draw [black , line width = 1.2pt ] (3,1.2) -- (3,0.8);
\draw [dashed ] (3,2) -- (3,0);

\node [above] at (4,1) { $q$};
\draw [blue, line width = 1.2pt  ] (3,1) -- (5,1);
\draw [blue, line width = 1.2pt  ] (5,1.2) -- (5,0.8);

\end{tikzpicture}
\caption{Interval $p$ meets interval $q$}
\label{fig:meets}
\end{figure}
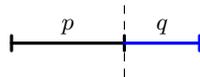

We consider the situations where two time intervals can be equal or meeting each other. Equality between interval is an equivalence relation. Two intervals meet if one interval ends at the starting time of the second.  The intervals are therefore  adjacent. For instance in Fig.~\ref{fig:meets}, interval $p$ meets interval $q$. The meets relation is irreflexive, non-symmetric and non-transitive. A set of five axioms (M1) $\sim$ (M5) is then defined based on equality and relation meets. 

We define a class type interval whose assumptions are (a) properties of relation meets, denoted by the infix symbol ``$||$" and, (b) axioms (M1) $\sim$ (M5). 
\begin{isabellebody}
\small\isanewline
\isacommand{class}\isamarkupfalse%
\ interval\ {\isacharequal}\isanewline
\ \isakeyword{fixes}\isanewline
\ \ meets{\isacharcolon}{\isacharcolon}{\isachardoublequoteopen}{\isacharprime}a\ {\isasymRightarrow}\ {\isacharprime}a\ {\isasymRightarrow}\ bool{\isachardoublequoteclose}\ \ {\isacharparenleft}\isakeyword{infixl}\ {\isachardoublequoteopen}{\isasymparallel}{\isachardoublequoteclose}\ {\isadigit{6}}{\isadigit{0}}{\isacharparenright}\ \isanewline
\ \isakeyword{assumes}\isanewline
\ \ meets{\isacharunderscore}atrans{\isacharcolon}{\isachardoublequoteopen}{\isasymlbrakk}p{\isasymparallel}q {\isacharsemicolon}  q{\isasymparallel}r{\isasymrbrakk}\ {\isasymLongrightarrow}\ {\isasymnot}{\isacharparenleft}p{\isasymparallel}r{\isacharparenright}{\isachardoublequoteclose}\ \isakeyword{and}\isanewline
\ \ meets{\isacharunderscore}irrefl{\isacharcolon}{\isachardoublequoteopen}{\isasymnot}{\isacharparenleft}p{\isasymparallel}p{\isacharparenright}{\isachardoublequoteclose}\ \isakeyword{and}\isanewline
\ \ meets{\isacharunderscore}asym{\isacharcolon}{\isachardoublequoteopen}p{\isasymparallel}q\ {\isasymLongrightarrow}\ {\isasymnot}{\isacharparenleft}q{\isasymparallel}p{\isacharparenright}{\isachardoublequoteclose}\ \isakeyword{and}\isanewline
\ \ M{\isadigit{1}}{\isacharcolon}{\isachardoublequoteopen}{\isasymlbrakk}p{\isasymparallel}q{\isacharsemicolon}\ p{\isasymparallel}s{\isacharsemicolon}\ r{\isasymparallel}q{\isasymrbrakk}\ {\isasymLongrightarrow}\ {\isacharparenleft}r{\isasymparallel}s{\isacharparenright}{\isachardoublequoteclose}\ \isakeyword{and}\isanewline
\ \ M{\isadigit{2}}{\isacharcolon}{\isachardoublequoteopen}{\isasymlbrakk}p{\isasymparallel}q {\isacharsemicolon}\  r{\isasymparallel}s{\isasymrbrakk}\ {\isasymLongrightarrow} \isanewline \ \ \ \ \ \ \ \ \ \ \ \  \ \ \ p{\isasymparallel}s\ {\isasymoplus}\ {\isacharparenleft}{\isacharparenleft}{\isasymexists}t{\isachardot}\ {\isacharparenleft}p{\isasymparallel}t{\isacharparenright}{\isasymand}{\isacharparenleft}t{\isasymparallel}s{\isacharparenright}{\isacharparenright}\ {\isasymoplus}\ {\isacharparenleft}{\isasymexists}t{\isachardot}\ {\isacharparenleft}r{\isasymparallel}t{\isacharparenright}{\isasymand}{\isacharparenleft}t{\isasymparallel}q{\isacharparenright}{\isacharparenright}{\isacharparenright}{\isachardoublequoteclose}\ \isakeyword{and}\isanewline
\ \ M{\isadigit{3}}{\isacharcolon}{\isachardoublequoteopen}{\isacharparenleft}{\isasymexists}q\ r{\isachardot}\ q{\isasymparallel}p\ {\isasymand}\ p{\isasymparallel}r{\isacharparenright}{\isachardoublequoteclose}\ \isakeyword{and}\isanewline
\ \ M{\isadigit{4}}{\isacharcolon}{\isachardoublequoteopen}{\isasymlbrakk}p{\isasymparallel}q\ {\isacharsemicolon}\ q{\isasymparallel}s\ {\isacharsemicolon}\ p{\isasymparallel}r\ {\isacharsemicolon}\ r{\isasymparallel}s{\isasymrbrakk}\ {\isasymLongrightarrow}\ q\ {\isacharequal}\ r{\isachardoublequoteclose}\ \ \isakeyword{and}\isanewline
\ \ M{\isadigit{5}}exist{\isacharcolon}{\isachardoublequoteopen}p{\isasymparallel}q\ {\isasymLongrightarrow}\ {\isacharparenleft}{\isasymexists}r\ s\ t{\isachardot}\ r{\isasymparallel}p\ {\isasymand}\ p{\isasymparallel}q\ {\isasymand}\ q{\isasymparallel}s\ {\isasymand}\ r{\isasymparallel}t\ {\isasymand}\ t{\isasymparallel}s{\isacharparenright}{\isachardoublequoteclose}\isanewline
\end{isabellebody}

\begin{figure}[h]
\scriptsize
\centering
\begin{tikzpicture}[x=.5cm,y=.5cm, scale =0.7]
\clip(-25,6) rectangle (10,23);

\draw [line width = 0.7pt] (0,20) -- (2,20);
\draw [line width = 0.7pt] (0,20.1) -- (0,19.9);
\draw [line width = 0.7pt] (2,20.1) -- (2,19.9);

\node [above] at (1,20) { $p$};
\draw [black, dashed ] (2,20.5) -- (2,18);
\draw [line width = 0.7pt, blue] (2,20) -- (4,20);
\draw [line width = 0.7pt, blue] (2,20.1) -- (2,19.9);
\draw [line width = 0.7pt, blue] (4,20.1) -- (4,19.9);
\node [above] at (3,20) { $q$};
\draw [line width = 0.7pt, blue] (2,19) -- (5,19);
\draw [line width = 0.7pt, blue] (2,19.1) -- (2,18.9);
\draw [line width = 0.7pt,blue] (5,19.1) -- (5,18.9);

\node [above] at (3.5,19) { $s$};

\draw [line width = 0.7pt] (0.7,18.5) -- (2,18.5);
\draw [line width = 0.7pt] (0.7,18.4) -- (0.7,18.6);
\draw [line width = 0.7pt] (2,18.4) -- (2,18.6);
\node [above] at (1.2,18.5) { $r$};

\node [right] at (-25, 19.4) {(M1) $\llbracket p||q;\; p||s ; \; r||q\rrbracket \Longrightarrow r||s$};
\draw[line width=0.2pt] (-25,17.3) -- (10,17.3);
\draw [line width = 0.7pt] (2,16) -- (4,16);

\node [above] at (3,16) { $p$};
\draw [black, dashed ] (2,16.5) -- (2,15.2);
\draw [black, dashed ] (4,16.5) -- (4,15.2);

\draw [line width = 0.7pt, red] (1,16) -- (2,16);
\draw [line width = 0.7pt, red] (2,16.1) -- (2,15.9);
\draw [line width = 0.7pt, red] (1,16.1) -- (1,15.9);

\node [above] at (1.5,16) { $q$};

\draw [line width = 0.7pt, red] (4,16) -- (6,16);
\draw [line width = 0.7pt, red] (4,16.1) -- (4,15.9);
\draw [line width = 0.7pt, red] (6,16.1) -- (6,15.9);

\node [above] at (5,16) { $r$};
\node [right] at (-25, 16) {(M3) $\exists q\; r. \; q||p \land p||r$};
\draw[line width = 0.2pt] (-25,14.8) -- (10,14.8);

\draw [line width = 0.7pt, blue] (2,13.5) -- (4,13.5);
\draw [line width = 0.7pt, blue] (2,13.6) -- (2,13.4);
\draw [line width = 0.7pt, blue] (4,13.6) -- (4,13.4);

\node [above] at (3,13.5) { $q$};
\draw [black, dashed ] (2,14.5) -- (2,11.2);
\draw [black, dashed ] (4,14.5) -- (4,11.2);

\draw [line width = 0.7pt] (1,13.5) -- (2,13.5);
\draw [line width = 0.7pt] (1,13.6) -- (1,13.4);

\node [above] at (1.5,13.5) { $p$};

\draw [line width = 0.7pt] (4,13.5) -- (6,13.5);
\draw [line width = 0.7pt] (6,13.6) -- (6,13.4);

\node [above] at (5,13.5) { $s$};

\draw [line width = 0.7pt, blue] (2,12) -- (4,12);
\draw [line width = 0.7pt, blue] (2,12.1) -- (2,11.9);
\draw [line width = 0.7pt, blue] (4,12.1) -- (4,11.9);

\node [above] at (3,12) { $r$};
\node [right] at (-25, 13) {(M4) $\llbracket q||p ;\; q||s; \; p||r ;\; r||s\rrbracket \Longrightarrow q = r$};
\draw[line width = 0.2pt] (-25,11.5) -- (10,11.5);

\draw [line width = 0.7pt] (2,9.5) -- (4,9.5);
\draw [line width = 0.7pt] (4,9.6) -- (4,9.4);

\node [above] at (3,9.5) { $p$};
\draw [line width = 0.7pt] (4,9.5) -- (5.5,9.5);
\node [above] at (4.8,9.5) { $q$};
\draw [black, dashed ] (2,7.5) -- (2,11);
\draw [black, dashed ] (4,7.5) -- (4,11);
\draw [black, dashed ] (5.5,7.5) -- (5.5,11);

\draw [line width = 0.7pt, blue] (0,9.5) -- (2,9.5);
\draw [line width = 0.7pt, blue] (0,9.6) -- (0,9.4);
\draw [line width = 0.7pt, blue] (2,9.6) -- (2,9.4);

\node [above] at (1,9.5) { $r$};

\draw [line width = 0.7pt, blue] (5.5,9.5) -- (7,9.5);
\draw [line width = 0.7pt, blue] (5.5,9.4) -- (5.5,9.6);
\draw [line width = 0.7pt, blue] (7,9.4) -- (7,9.6);

\node [above] at (6.3,9.5) { $s$};

\draw [line width = 0.7pt, red] (2,8) -- (5.5,8);
\draw [line width = 0.7pt, red] (2,8.9) -- (2,8.1);
\draw [line width = 0.7pt, red] (5.5,8.9) -- (5.5,8.1);

\node [above] at (4,8) { $t$};
\node [right] at (-25, 10) {(M5) $ p||q \Longrightarrow (\exists r\;s\;t. \;r||p\land p||q \land q||s \land r||t \land t||s)$};
\end{tikzpicture}
\caption{Axioms (M1), (M3) $\sim$ (M5)}
\label{fig:M1M5}
\end{figure}
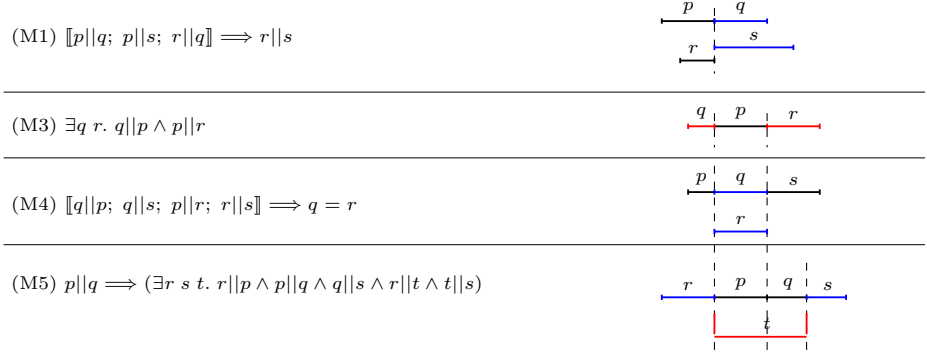
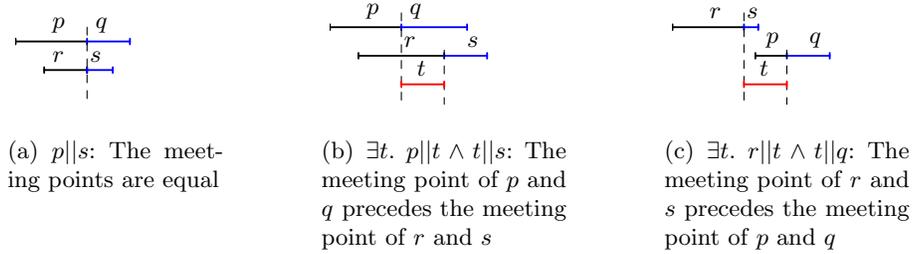
\begin{figure}
\centering
\subfloat[$p||s$: The meeting points are equal]{
\begin{tikzpicture}[line cap=round,line join=round,>=triangle 45,x=1.0cm,y=1.0cm, scale = 0.38]
\clip(-1,-2) rectangle (6,2);
\draw [line width = 0.7pt] (-1,1) -- (1.5,1);
\draw [line width = 0.7pt] (-1,1.1) -- (-1,0.9);

\node [above] at (0.5,1) { $p$};
\draw [line width = 0.7pt, blue] (1.5,1) -- (3,1);
\draw [line width = 0.7pt,blue] (1.5,1.1) -- (1.5,0.9);
\draw [line width = 0.7pt,blue] (3,1.1) -- (3,0.9);

\node [above] at (2,1) { $q$};
\draw [black, dashed ] (1.5,1.5) -- (1.5,-1.2);

\draw [line width = 0.7pt] (0,0) -- (1.5,0);
\draw [line width = 0.7pt] (0,0.1) -- (0,-0.1);
\draw [line width = 0.7pt] (1.5,0.1) -- (1.5,-0.1);

\node [above] at (0.5,0) { $r$};

\draw [line width = 0.7pt, blue] (1.5,0) -- (2.4,0);
\draw [line width = 0.7pt, blue] (1.5,0.1) -- (1.5,-0.1);
\draw [line width = 0.7pt, blue] (2.4,0.1) -- (2.4,-0.1);

\node [above] at (1.8,0) { $s$};

\end{tikzpicture}\label{fig:M2c1}
} \qquad\qquad
\subfloat[$\exists t.~p||t \land t ||s$: The meeting point of $p$ and $q$ precedes the meeting point of $r$ and $s$]{
\begin{tikzpicture}[line cap=round,line join=round,>=triangle 45,x=1.0cm,y=1.0cm, scale = 0.38]
\clip(-1,-2) rectangle (7,2.5);
\draw [line width = 0.7pt] (-1,1.5) -- (1.5,1.5);
\draw [line width = 0.7pt] (-1,1.4) -- (-1,1.6);

\node [above] at (0.5,1.5) { $p$};
\draw [line width = 0.7pt, blue] (1.5,1.5) -- (3.8,1.5);
\draw [line width = 0.7pt,blue] (1.5,1.4) -- (1.5,1.6);
\draw [line width = 0.7pt,blue] (3.8,1.4) -- (3.8,1.6);

\node [above] at (2,1.5) { $q$};
\draw [black, dashed ] (1.5,2) -- (1.5,-1.2);

\draw [line width = 0.7pt] (0,0.5) -- (3,0.5);
\draw [line width = 0.7pt] (0,0.4) -- (0,0.6);

\node [above] at (1.8,0.5) { $r$};

\draw [line width = 0.7pt, blue] (3,0.5) -- (4.5,0.5);
\draw [line width = 0.7pt, blue] (3,0.4) -- (3,0.6);
\draw [line width = 0.7pt, blue] (4.5,0.4) -- (4.5,0.6);

\node [above] at (4,0.5) { $s$};
\draw [black, dashed ] (3,0.7) -- (3,-1.2);

\draw [line width = 0.7pt, red] (1.5,-0.5) -- (3,-0.5);
\draw [line width = 0.7pt, red] (1.5,-0.4) -- (1.5,-0.6);
\draw [line width = 0.7pt, red] (3,-0.4) -- (3,-0.6);

\node [above] at (2.2,-0.5) { $t$};

\end{tikzpicture}\label{fig:M2c2}
} \qquad\qquad
\subfloat[$\exists t.~r||t\land t||q$: The meeting point of $r$ and $s$ precedes the meeting point of $p$ and $q$]{
\begin{tikzpicture}[line cap=round,line join=round,>=triangle 45,x=1.0cm,y=1.0cm, scale = 0.38]
\clip(-1,-2) rectangle (7,2.5);
\draw [line width = 0.7pt] (-1,1.5) -- (1.5,1.5);
\draw [line width = 0.7pt] (-1,1.4) -- (-1,1.6);

\node [above] at (0.5,1.5) { $r$};
\draw [line width = 0.7pt, blue] (1.5,1.5) -- (2,1.5);
\draw [line width = 0.7pt,blue] (1.5,1.4) -- (1.5,1.6);
\draw [line width = 0.7pt,blue] (2,1.4) -- (2,1.6);

\node [above] at (1.8,1.5) { $s$};
\draw [black, dashed ] (1.5,2) -- (1.5,-1.2);

\draw [line width = 0.7pt] (1.9,0.5) -- (3,0.5);
\draw [line width = 0.7pt] (1.9,0.4) -- (1.9,0.6);

\node [above] at (2.5,0.5) { $p$};

\draw [line width = 0.7pt, blue] (3,0.5) -- (4.5,0.5);
\draw [line width = 0.7pt,blue] (3,0.4) -- (3,0.6);
\draw [line width = 0.7pt,blue] (4.5,0.4) -- (4.5,0.6);

\node [above] at (4,0.5) { $q$};
\draw [black, dashed ] (3,0.7) -- (3,-1.2);

\draw [line width = 0.7pt, red] (1.5,-0.5) -- (3,-0.5);
\draw [line width = 0.7pt, red] (1.5,-0.4) -- (1.5,-0.6);
\draw [line width = 0.7pt, red] (3,-0.4) -- (3,-0.6);

\node [above] at (2.2,-0.5) { $t$};
\end{tikzpicture}\label{fig:M2c3}
}

\caption{The three cases of axiom (M2)}
\label{fig:M2}
\end{figure}

Figures \ref{fig:M1M5} and \ref{fig:M2} illustrate the five time interval axioms:
\begin{itemize}
\item Axiom (M1) uniquely defines the meeting point of two intervals. If two intervals $p$ and $r$ meet the same interval, then $r$ meets any interval that $p$ meets. 
\item Axiom (M3) states that for any interval $p$ there is an interval that meets $p$ and an interval that it meets, which means there a time interval does not last infinitly.
\item Axiom (M4) states that an interval is uniquely defined by the starting and the ending point. 
\item Axiom (M5) states that the addition of two intervals that meet is an intervals. 
\item Axiom (M2) takes two pairs of intervals that meet in two points, and states that  either the meeting points are equal or one of them precedes the other.  Axiom (M2) states that the three configurations are mutually exclusive, i.e. not more than one can hold, and exhaustive, i.e. at least one of the three must hold.
\end{itemize}

\section{Formalization of Time Interval Relations}
\label{sect:arel}

A basic relation is a set of interval pairs  and of type {\isacharparenleft}{\isacharprime}a{\isasymtimes}{\isacharprime}a{\isacharparenright}\ set.  We extend type class \texttt{interval} with basic relations. 
\begin{isabellebody}
\isanewline
\isacommand{class}\isamarkupfalse%
\ arelations\ {\isacharequal}\ interval\ {\isacharplus}\ \isanewline
\ \isakeyword{fixes}\ \isanewline
\ \ e{\isacharcolon}{\isacharcolon}{\isachardoublequoteopen}{\isacharparenleft}{\isacharprime}a{\isasymtimes}{\isacharprime}a{\isacharparenright}\ set{\isachardoublequoteclose}\ \isakeyword{and}\isanewline
\ \ m{\isacharcolon}{\isacharcolon}{\isachardoublequoteopen}{\isacharparenleft}{\isacharprime}a{\isasymtimes}{\isacharprime}a{\isacharparenright}\ set{\isachardoublequoteclose}\ \isakeyword{and}\ \isanewline
\ \ b{\isacharcolon}{\isacharcolon}{\isachardoublequoteopen}{\isacharparenleft}{\isacharprime}a{\isasymtimes}{\isacharprime}a{\isacharparenright}\ set{\isachardoublequoteclose}\ \isakeyword{and}\isanewline
\ \ ov{\isacharcolon}{\isacharcolon}{\isachardoublequoteopen}{\isacharparenleft}{\isacharprime}a{\isasymtimes}{\isacharprime}a{\isacharparenright}\ set{\isachardoublequoteclose}\ \isakeyword{and}\isanewline
\ \ d{\isacharcolon}{\isacharcolon}{\isachardoublequoteopen}{\isacharparenleft}{\isacharprime}a{\isasymtimes}{\isacharprime}a{\isacharparenright}\ set{\isachardoublequoteclose}\ \isakeyword{and}\isanewline
\ \ s{\isacharcolon}{\isacharcolon}{\isachardoublequoteopen}{\isacharparenleft}{\isacharprime}a{\isasymtimes}{\isacharprime}a{\isacharparenright}\ set{\isachardoublequoteclose}\ \isakeyword{and}\isanewline
\ \ f{\isacharcolon}{\isacharcolon}{\isachardoublequoteopen}{\isacharparenleft}{\isacharprime}a{\isasymtimes}{\isacharprime}a{\isacharparenright}\ set{\isachardoublequoteclose}\ \ \isanewline
\isakeyword{assumes}\isanewline
\ \ e{\isacharcolon}{\isachardoublequoteopen}{\isacharparenleft}p{\isacharcomma}q{\isacharparenright}\ {\isasymin}\ e\ {\isacharequal}\ {\isacharparenleft}p\ {\isacharequal}\ q{\isacharparenright}{\isachardoublequoteclose}\ \isakeyword{and}\isanewline
\ \ m{\isacharcolon}{\isachardoublequoteopen}{\isacharparenleft}p{\isacharcomma}q{\isacharparenright}\ {\isasymin}\ m\ {\isacharequal}\ p{\isasymparallel}q{\isachardoublequoteclose}\ \isakeyword{and}\isanewline
\ \ b{\isacharcolon}{\isachardoublequoteopen}{\isacharparenleft}p{\isacharcomma}q{\isacharparenright}\ {\isasymin}\ b\ {\isacharequal}\ {\isacharparenleft}{\isasymexists}t{\isacharcolon}{\isacharcolon}{\isacharprime}a{\isachardot}\ p{\isasymparallel}t\ {\isasymand}\ t{\isasymparallel}q{\isacharparenright}{\isachardoublequoteclose}\ \isakeyword{and}\isanewline
\ \ ov{\isacharcolon}{\isachardoublequoteopen}{\isacharparenleft}p{\isacharcomma}q{\isacharparenright}\ {\isasymin}\ ov\ {\isacharequal}\ {\isacharparenleft}{\isasymexists}k\ l\ u\ v\ t{\isacharcolon}{\isacharcolon}{\isacharprime}a{\isachardot}\ {\isacharparenleft}k{\isasymparallel}p\ {\isasymand}\ p{\isasymparallel}u\ {\isasymand}\ u{\isasymparallel}v{\isacharparenright}\ {\isasymand}\   \isanewline
\ \ \ \ \ \ \ \ \ \ \ \ \ \ \ \ \ \ \ {\isacharparenleft}k{\isasymparallel}l\ {\isasymand}\ l{\isasymparallel}q\ {\isasymand}\ q{\isasymparallel}v{\isacharparenright}\ {\isasymand}\ {\isacharparenleft}l{\isasymparallel}t\ {\isasymand}\ t{\isasymparallel}u{\isacharparenright}{\isacharparenright}{\isachardoublequoteclose}\ \isakeyword{and}\isanewline
\ \ s{\isacharcolon}{\isachardoublequoteopen}{\isacharparenleft}p{\isacharcomma}q{\isacharparenright}\ {\isasymin}\ s\ {\isacharequal}\ \ {\isacharparenleft}{\isasymexists}k\ u\ v{\isacharcolon}{\isacharcolon}{\isacharprime}a{\isachardot}\ k{\isasymparallel}p\ {\isasymand}\ p{\isasymparallel}u\ {\isasymand}\ u{\isasymparallel}v\ {\isasymand}\ k{\isasymparallel}q\ {\isasymand}\ q{\isasymparallel}v{\isacharparenright}{\isachardoublequoteclose}\ \isakeyword{and}\isanewline
\ \ f{\isacharcolon}{\isachardoublequoteopen}{\isacharparenleft}p{\isacharcomma}q{\isacharparenright}\ {\isasymin}\ f\ {\isacharequal}\ {\isacharparenleft}{\isasymexists}k\ l\  u {\isacharcolon}{\isacharcolon}{\isacharprime}a{\isachardot}\ k{\isasymparallel}l\ {\isasymand}\ l{\isasymparallel}p\ {\isasymand}\ p{\isasymparallel}u\ {\isasymand}\ k{\isasymparallel}q\ {\isasymand}\ q{\isasymparallel}u{\isacharparenright}{\isachardoublequoteclose}\ \isakeyword{and}\isanewline
\ \ d{\isacharcolon}{\isachardoublequoteopen}{\isacharparenleft}p{\isacharcomma}q{\isacharparenright}\ {\isasymin}\ d\ {\isacharequal}\ {\isacharparenleft}{\isasymexists}k\ l\ u\ v{\isacharcolon}{\isacharcolon}{\isacharprime}a{\isachardot}\ k{\isasymparallel}l\ {\isasymand}\ l{\isasymparallel}p\ {\isasymand}\ p{\isasymparallel}u\ {\isasymand}u{\isasymparallel}v\ {\isasymand}\ k{\isasymparallel}q\ {\isasymand}\ q{\isasymparallel}v{\isacharparenright}{\isachardoublequoteclose}\isanewline %
\end{isabellebody}

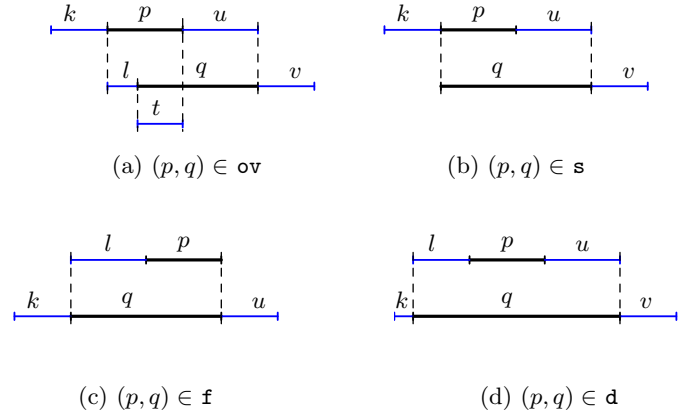
\begin{figure}
\centering
\subfloat[$(p,q)\in$ \texttt{ov}]
{\begin{tikzpicture}[line cap=round,line join=round,>=triangle 45,x=1.0cm,y=1.0cm, scale = 0.5]
\clip(-1,-2) rectangle (7,1.8);
\node [above] at (2,1) { $p$};
\draw [line width = 1.2pt] (1,1) -- (3,1);
\draw [line width = 0.7pt] (1,1.1) -- (1,0.9);
\draw [line width = 0.7pt] (3,1.1) -- (3,0.9);

\node [above] at (3.5,-0.5) { $q$};
\draw [line width = 1.2pt] (1.8,-0.5) -- (5,-0.5);
\draw [line width = 0.7pt] (1.8,-0.4) -- (1.8,-0.6);
\draw [line width = 0.7pt] (5,-0.4) -- (5,-0.6);

\node [above] at (0,1) {$k$};
\draw [line width = 0.7pt, blue] (-0.5,1) -- (1,1);
\draw [line width = 0.7pt, blue] (-0.5,1.1) -- (-0.5,0.9);
\node [above] at (4,1) {$u$};
\draw [line width = 0.7pt, blue] (3,1) -- (5,1);
\draw [line width = 0.7pt, blue] (5,1.1) -- (5,0.9);
\node [above] at (1.5,-0.5) {$l$};
\draw [line width = 0.7pt, blue] (1,-0.5) -- (1.8,-0.5);
\draw [line width = 0.7pt, blue] (1,-0.4) -- (1,-0.6);
\node [above] at (6,-0.5) {$v$};
\draw [line width = 0.7pt, blue] (5,-0.5) -- (6.5,-0.5);
\draw [line width = 0.7pt, blue] (6.5,-0.4) -- (6.5,-0.6);
\node [above] at (2.3,-1.5) { $t$};
\draw [line width = 0.7pt,blue] (1.8,-1.5) -- (3,-1.5);
\draw [line width = 0.7pt,blue] (1.8,-1.4) -- (1.8,-1.6);
\draw [line width = 0.7pt,blue] (3,-1.4) -- (3,-1.6);

\draw [line width=0.5pt, dashed] (1,1.2) -- (1,-0.7);
\draw [line width=0.5pt, dashed] (5,1.2) -- (5,-0.7);
\draw [line width=0.5pt, dashed] (1.8,-0.3) -- (1.8,-1.7);
\draw [line width=0.5pt, dashed] (3,1.3) -- (3,-1.7);
\end{tikzpicture}\label{fig:ov}
} \quad
\subfloat[$(p,q)\in$ \texttt{s}]
{\begin{tikzpicture}[line cap=round,line join=round,>=triangle 45,x=1.0cm,y=1.0cm, scale = 0.5]
\clip(-1,-2) rectangle (7,3);
\node [above] at (2,1) { $p$};
\draw [line width = 1.2pt] (1,1) -- (3,1);
\draw [line width = 0.7pt] (1,1.1) -- (1,0.9);
\draw [line width = 0.7pt] (3,1.1) -- (3,0.9);

\node [above] at (2.5,-0.5) { $q$};
\draw [line width = 1.2pt] (1,-0.5) -- (5,-0.5);
\draw [line width = 0.7pt] (1,-0.4) -- (1,-0.6);
\draw [line width = 0.7pt] (5,-0.4) -- (5,-0.6);

\node [above] at (0,1) {$k$};
\draw [line width = 0.7pt, blue] (-0.5,1) -- (1,1);
\draw [line width = 0.7pt, blue] (-0.5,1.1) -- (-0.5,0.9);

\node [above] at (4,1) {$u$};
\draw [line width = 0.7pt, blue] (3,1) -- (5,1);
\draw [line width = 0.7pt, blue] (5,1.1) -- (5,0.9);

\node [above] at (6,-0.5) {$v$};
\draw [line width = 0.7pt, blue] (5,-0.5) -- (6.5,-0.5);
\draw [line width = 0.7pt, blue] (6.5,-0.4) -- (6.5,-0.6);

\draw [line width=0.5pt, dashed] (1,1.2) -- (1,-0.7);
\draw [line width=0.5pt, dashed] (5,1.2) -- (5,-0.7);

\end{tikzpicture}\label{fig:s}}\qquad\qquad
\subfloat[$(p,q)\in$ \texttt{f}]
{\begin{tikzpicture}[line cap=round,line join=round,>=triangle 45,x=1.0cm,y=1.0cm, scale = 0.5]
\clip(-1,-2) rectangle (7,3);
\node [above] at (4,1) { $p$};
\draw [line width = 1.2pt] (3,1) -- (5,1);
\draw [line width = 0.7pt] (3,1.1) -- (3,0.9);
\draw [line width = 0.7pt] (5,1.1) -- (5,0.9);

\node [above] at (2.5,-0.5) { $q$};
\draw [line width = 1.2pt] (1,-0.5) -- (5,-0.5);
\draw [line width = 0.7pt] (1,-0.4) -- (1,-0.6);
\draw [line width = 0.7pt] (5,-0.4) -- (5,-0.6);

\node [above] at (0,-0.5) {$k$};
\draw [line width = 0.7pt, blue] (-0.5,-0.5) -- (1,-0.5);
\draw [line width = 0.7pt, blue] (-0.5,-0.4) -- (-0.5,-0.6);

\node [above] at (2,1) {$l$};
\draw [line width = 0.7pt, blue] (1,1) -- (3,1);
\draw [line width = 0.7pt, blue] (1,1.1) -- (1,0.9);
\node [above] at (6,-0.5) {$u$};
\draw [line width = 0.7pt, blue] (5,-0.5) -- (6.5,-0.5);
\draw [line width = 0.7pt, blue] (6.5,-0.4) -- (6.5,-0.6);
\draw [line width=0.5pt, dashed] (1,1.2) -- (1,-0.7);
\draw [line width=0.5pt, dashed] (5,1.2) -- (5,-0.7);

\end{tikzpicture}\label{fig:f}}\qquad\qquad
\subfloat[$(p,q)\in$ \texttt{d}]
{\begin{tikzpicture}[line cap=round,line join=round,>=triangle 45,x=1.0cm,y=1.0cm, scale = 0.5]
\clip(-1,-2) rectangle (7,3);

\node [above] at (2,1) { $p$};
\draw [line width = 1.2pt] (1,1) -- (3,1);
\draw [line width = 0.7pt] (1,1.1) -- (1,0.9);
\draw [line width = 0.7pt] (3,1.1) -- (3,0.9);

\node [above] at (2,-0.5) { $q$};
\draw [line width = 1.2pt] (-0.5,-0.5) -- (5,-0.5);
\draw [line width = 0.7pt] (-0.5,-0.4) -- (-0.5,-0.6);
\draw [line width = 0.7pt] (5,-0.4) -- (5,-0.6);

\node [above] at (-0.8,-0.5) {$k$};
\draw [line width = 0.7pt, blue] (-1,-0.5) -- (-0.5,-0.5);
\draw [line width = 0.7pt, blue] (-1,-0.4) -- (-1,-0.6);
\node [above] at (0,1) {$l$};
\draw [line width = 0.7pt, blue] (-0.5,1) -- (1,1);
\draw [line width = 0.7pt, blue] (-0.5,1.1) -- (-0.5,0.9);

\node [above] at (4,1) {$u$};
\draw [line width = 0.7pt, blue] (3,1) -- (5,1);
\draw [line width = 0.7pt, blue] (5,1.1) -- (5,0.9);

\node [above] at (5.7,-0.5) {$v$};
\draw [line width = 0.7pt, blue] (5,-0.5) -- (6.5,-0.5);
\draw [line width = 0.7pt, blue] (6.5,-0.4) -- (6.5,-0.6);

\draw [line width=0.5pt, dashed] (-0.5,1.2) -- (-0.5,-0.7);
\draw [line width=0.5pt, dashed] (5,1.2) -- (5,-0.7);

\end{tikzpicture}\label{fig:d}
} 
\caption{Relations \texttt{ov}, \texttt{s}, \texttt{f} and \texttt{d} defined as conjunctions of literals of the form $x||y$, where $x$ and $y$ are time intervals}
\label{fig:arelations}
\end{figure}

In the assumptions of class \texttt{arelations}, each basic time interval relation is rewritten as a first order formula in a prenex normal form, i.e.~a conjunction of atomic formulas $x || y$ and $x  = y$. 
Relations \texttt{e}, \texttt{m} and \texttt{b}  are self-explanatory. Rewriting rules of the remaining relations are illustrated in Fig.~\ref{fig:arelations}. 

We call literal the atomic formula of the form $x||y$ and $x=y$. 
We denote by $L_r(p,q)$ the set of literals generated by the relation membership $(p,q) \in r$. Among the literals in $L_r(p,q)$, there are constraints on the starting points and the ending points of intervals $p$ and $q$. For instance,  $(p,q) \in$ \texttt{ov} gives rise to the eight literals \{$k||p$, $p||u$, $u||v$, $k||l$, $l||q$, $q||v$, $l||t$ , $t||u$\} = $L_{\texttt{ov}}(p,q)$. Literals $k||p$ and $l||q$ are constraints on the starting points of $p$ and $q$. Literals $p||u$ and $q||v$ are constraints on the ending points of $p$ and $q$.

%

\section{JEPD Property}
\label{sect:jepd}
In qualitative reasoning, representations of events or objects in the space are based on a set of relations that are jointly exhaustive (JE) and pairwise disjoint (PD). JE property ensures the expressiveness of the relations.  
In other words, for any two intervals, there exists a relation that holds between these two intervals. The 13 time interval relations, therefore, constitute a partition of the set of interval pairs. 

The unicity of a relation between two arbitrary intervals is allowed by the PD property.  Therefore, the 13 relations enable us to express precise information between any two temporal intervals and avoid redundancy. 

\subsubsection{PD property.} 
For any two different basic relations $r_1$ and $r_2$, we show that $r_1\; \cap\; r_2 = \{\}$. The proof is performed by contradiction. From $(p,q) \in r_1$ and $(p,q) \in r_2$, we deduce literals that refute the non-transitivity property of ``$||$", i.e. property \texttt{meets\_atrans}. 
\begin{isabellebody}
\isanewline
\isacommand{lemma}\isamarkupfalse%
\ {\isachardoublequoteopen}m\ {\isasyminter}\ b\ {\isacharequal}\ {\isacharbraceleft}{\isacharbraceright}{\isachardoublequoteclose}\isanewline
%
%
%
\isacommand{using}\isamarkupfalse%
\ m\ b\ \isacommand{apply}\isamarkupfalse%
\ auto\isanewline
\isacommand{using}\isamarkupfalse%
\ meets\_atrans\ \isacommand{by}\isamarkupfalse%
\ blast%
\isanewline
{\isafoldproof}%
%
\end{isabellebody}


Relations $r_1$ and $r_2$ sometimes generate literals from which we cannot immediately deduce the contradiction. Therefore, we use the axiomatic system defined in Sect.~\ref{sect:axioms} to deduce new ``$||$" relations and obtain new intervals. The tactics \texttt{metis} and \texttt{meson}  finish the proof successfully. 
\begin{isabellebody}
\isanewline
\isacommand{lemma}\isamarkupfalse%
 \ {\isachardoublequoteopen}s\ {\isasyminter}\ d\ {\isacharequal}\ {\isacharbraceleft}{\isacharbraceright}{\isachardoublequoteclose}\ \isanewline
%
%
\isacommand{apply}\isamarkupfalse%
\ \isacommand{using}\isamarkupfalse%
\ s\ d\ \isacommand{apply}\isamarkupfalse%
\ auto\isanewline
\isacommand{by}\isamarkupfalse%
\ {\isacharparenleft}meson\ M{\isadigit{1}}\ meets{\isacharunderscore}atrans{\isacharparenright}%
{\isafoldproof}%
%
\isanewline
\end{isabellebody}


\subsubsection{JE property.} 
Let $\delta$ be the relation that is  the union of all basic relations. $$\delta = (\texttt{b},\texttt{m},\texttt{ov}, \texttt{s}, \texttt{d}, \texttt{f}, \texttt{e}, \texttt{f}^{-1}, \texttt{d}^{-1}, \texttt{s}^{-1}, \texttt{ov}^{-1}, \texttt{m}^{-1}, \texttt{b}^{-1}).$$ 
JE property means that for any two intervals $x$ and $y$, the expression $(x, y) \in \delta$ always holds. We do not explain the proof as it is similar to the proof of $\delta$-composition in Sect.~\ref{sect:gammadelta}.

%
\section{Compositions of Time Interval Relations}
\label{sect:comp}
\subsection{Composition Table}
The compositions of time interval relations were first computed by Allen~\cite{Allen:1985}. We arrange the result of the compositions in Table~\ref{table1}. We read the table as follows. Let $r_1$ and $r_2$ be two of the relations given in the first column and the first row, respectively. An entry in the table is a relation that contains the composition $r_1\circ r_2$. For instance, according to the table, the composition \texttt{s} $\circ$ \texttt{m}  is a subset of relation \texttt{b}. We write \texttt{s} $\circ$ \texttt{m} $\subseteq$ \texttt{b}. Some compositions give rise to union of basic relations, e.g.~\texttt{b} $\circ$ \texttt{d}. We call them relations $\alpha$, $\beta$, $\gamma$ and $\delta$ and define them as follows.
\begin{align*}
& \alpha_1 = (\texttt{ov}, \texttt{s}, \texttt{d}) \\ 
& \alpha_2 = (\texttt{ov}, \texttt{f}^{-1}, \texttt{d}^{-1}) \\ 
& \alpha_3 = (\texttt{b}, \texttt{m}, \texttt{ov}) \\ 
&\alpha_4 = (\texttt{f}^{-1}, \texttt{e}, \texttt{f})\\
& \alpha_5 =  (\texttt{s}, \texttt{e}, \texttt{s}^{-1})\\
& \beta_1 = (\texttt{b}, \texttt{m}, \texttt{ov}, \texttt{s}, \texttt{d}) \\ 
& \beta_2 = (\texttt{b}, \texttt{m}, \texttt{ov}, \texttt{f}^{-1}, \texttt{d}^{-1})\\ 
& \gamma = (\texttt{ov}, \texttt{s}, \texttt{d}, \texttt{f}, \texttt{e}, \texttt{f}^{-1}, \texttt{d}^{-1}, \texttt{s}^{-1}, \texttt{ov}^{-1})\\
& \delta = (\texttt{b},\texttt{m},\texttt{ov}, \texttt{s}, \texttt{d}, \texttt{f}, \texttt{e}, \texttt{f}^{-1}, \texttt{d}^{-1}, \texttt{s}^{-1}, \texttt{ov}^{-1}, \texttt{m}^{-1}, \texttt{b}^{-1})
\end{align*}

Their inverses are deduced from the fact that the inverse of a union of relations is a union of inverse relations. 
\begin{align*}
& {\alpha_1}^{-1} = (\texttt{ov}^{-1}, \texttt{s}^{-1}, \texttt{d}^{-1}) \\
& {\alpha_2}^{-1} = (\texttt{ov}^{-1}, \texttt{f}, \texttt{d})\\
& {\alpha_3}^{-1} = (\texttt{b}^{-1}, \texttt{m}^{-1}, \texttt{ov}^{-1})\\
& {\alpha_4}^{-1} = (\texttt{f}^{-1}, \texttt{e}, \texttt{f})\\
&  {\alpha_5}^{-1} = (\texttt{s}, \texttt{e}, \texttt{s}^{-1})\\
&  {\beta_1}^{-1} = (\texttt{b}^{-1}, \texttt{m}^{-1}, \texttt{ov}^{-1}, \texttt{s}^{-1}, \texttt{d}^{-1})\\  
& {\beta_2}^{-1} = (\texttt{b}^{-1}, \texttt{m}^{-1}, \texttt{ov}^{-1}, \texttt{f}, \texttt{d})\\ 
& {\gamma}^{-1} = (\texttt{ov}, \texttt{s}, \texttt{d}, \texttt{f}, \texttt{e}, \texttt{f}^{-1}, \texttt{d}^{-1}, \texttt{s}^{-1}, \texttt{ov}^{-1})\\
& {\delta}^{-1} = (\texttt{b},\texttt{m},\texttt{ov}, \texttt{s}, \texttt{d}, \texttt{f}, \texttt{e}, \texttt{f}^{-1}, \texttt{d}^{-1}, \texttt{s}^{-1}, \texttt{ov}^{-1}, \texttt{m}^{-1}, \texttt{b}^{-1})
\end{align*}

The Table \ref{table1} is symmetric w.r.t.~the diagonal white cells. The relations in yellow cells  are inverse of the relations in blue cells. For instance, referring to the table, we have $\texttt{s}\; \circ \;\texttt{d}^{-1} \subseteq \beta_2$. By symmetry w.r.t.~the diagonal, we reach the yellow cell that corresponds to $\texttt{d}\; \circ\; \texttt{s}^{-1} = (\texttt{s}\; \circ \texttt{d}^{-1})^{-1} \subseteq {\beta_2}^{-1}$. Relations \texttt{e}, $\alpha_4$, $\alpha_5$, $\gamma$ and $\delta$ are closed under inversion and they appear only in the diagonal white cells. What we need to prove is the compositions that correspond to blue and white cells, the proofs of compositions given by yellow cells are then deduced immediately. 

\begin{table*}[h]
\begin{center}
{\caption{The composition table of  time interval relations}\label{table1}}
\small
\begin{tabular}{|c|p{0.7cm}|p{0.7cm}|p{0.7cm}|p{0.7cm}|p{0.7cm}|p{0.7cm}|p{0.7cm}|p{0.7cm}|p{0.7cm}|p{0.7cm}|p{0.7cm}|p{0.7cm}|p{0.7cm}|}
\hline
\cellcolor{gray!8}\backslashbox{$r_1$}{$r_2$} & \cellcolor{gray!8}$b$ & \cellcolor{gray!8} $m$ & \cellcolor{gray!8}$ov$ & \cellcolor{gray!8}$f^{-1}$ & \cellcolor{gray!8}$d^{-1}$ & \cellcolor{gray!8}$s$ & \cellcolor{gray!8}$e$ & \cellcolor{gray!8}$s^{-1}$ & \cellcolor{gray!8}$d$ & \cellcolor{gray!8}$f$ & \cellcolor{gray!8}$ov^{-1}$ & \cellcolor{gray!8}$m^{-1}$ & \cellcolor{gray!8}$b^{-1}$ \\
\hline
\cellcolor{gray!8}$b$ & \cellcolor{blue!20}$b$  & \cellcolor{blue!20}$b$  & \cellcolor{blue!20}$b$  & \cellcolor{blue!20}$b$  & \cellcolor{blue!20}$b$  & \cellcolor{blue!20}$b$  & \cellcolor{blue!20}$b$  & \cellcolor{blue!20}$b$  & \cellcolor{blue!20}$\beta_1$ & \cellcolor{blue!20}$\beta_1$ & \cellcolor{blue!20}$\beta_1$ & \cellcolor{blue!20}$\beta_1$  & $\delta$ \\
\hline
\cellcolor{gray!8}$m$ & \cellcolor{blue!20}$b$ & \cellcolor{blue!20}$b$ &\cellcolor{blue!20} $b$ & \cellcolor{blue!20}$b$ & \cellcolor{blue!20}$b$  & \cellcolor{blue!20}$m$& \cellcolor{blue!20}$m$& \cellcolor{blue!20}$m$ & \cellcolor{blue!20}$\alpha_1$ & \cellcolor{blue!20}$\alpha_1$& \cellcolor{blue!20}$\alpha_1$ & $\alpha_4$ & \cellcolor{yellow!30}${\beta_1}^{-1}$ \\
\hline
\cellcolor{gray!8}$ov$ & \cellcolor{blue!20}$b$ & \cellcolor{blue!20}$b$ & \cellcolor{blue!20}$\alpha_3$ & \cellcolor{blue!20}$\alpha_3$ & \cellcolor{blue!20}$\beta_2$ & \cellcolor{blue!20}$ov$ & \cellcolor{blue!20}$ov$ & \cellcolor{blue!20}$\alpha_2$ & \cellcolor{blue!20}$\alpha_1$ & \cellcolor{blue!20}$\alpha_1$ & $\gamma$ & \cellcolor{yellow!30}${\alpha_1}^{-1}$ & \cellcolor{yellow!30}${\beta_1}^{-1}$ \\
\hline
\cellcolor{gray!8}$f^{-1}$ & \cellcolor{blue!20}$b$ & \cellcolor{blue!20}$m$ & \cellcolor{blue!20}$ov$ & \cellcolor{blue!20}$f^{-1}$ & \cellcolor{blue!20}$d^{-1}$ & \cellcolor{blue!20}$ov$ & \cellcolor{blue!20}$f^{-1}$ & \cellcolor{blue!20}$d^{-1}$ & \cellcolor{blue!20}$\alpha_1$ & $\alpha_4$ & \cellcolor{yellow!30}$\alpha_1^{-1}$ & \cellcolor{yellow!30}$\alpha_1^{-1}$ & \cellcolor{yellow!30}$\beta_1^{-1}$ \\
\hline
\cellcolor{gray!8}$d^{-1}$ & \cellcolor{blue!20}$\beta_2$ & \cellcolor{blue!20}$\alpha_2$ & \cellcolor{blue!20}$\alpha_2$ & \cellcolor{blue!20}$d^{-1}$ & \cellcolor{blue!20}$d^{-1}$ & \cellcolor{blue!20}$\alpha_2$ & \cellcolor{blue!20}$d^{-1}$ & \cellcolor{blue!20}$d^{-1}$ & $\gamma$ & \cellcolor{yellow!30}$\alpha_1^{-1}$ & \cellcolor{yellow!30}$\alpha_1^{-1}$ & \cellcolor{yellow!30}$\alpha_1^{-1}$ &  \cellcolor{yellow!30}$\beta_1^{-1}$ \\
\hline
\cellcolor{gray!8}$s$ & \cellcolor{blue!20}$b$ & \cellcolor{blue!20}$b$ & \cellcolor{blue!20}$\alpha_3$ & \cellcolor{blue!20}$\alpha_3$ & {\cellcolor{blue!20}$\beta_2$} & \cellcolor{blue!20}$s$ & \cellcolor{blue!20}$s$ & $\alpha_5$ & \cellcolor{yellow!30}$d$ & \cellcolor{yellow!30}$d$ & \cellcolor{yellow!30}$\alpha_2^{-1}$ & \cellcolor{yellow!30}$m^{-1}$ &  \cellcolor{yellow!30}$b^{-1}$ \\
\hline
\cellcolor{gray!8}$e$ & \cellcolor{blue!20}$b$ & \cellcolor{blue!20}$m$ & \cellcolor{blue!20}$ov$ & \cellcolor{blue!20}$f^{-1}$ & \cellcolor{blue!20}$d^{-1}$ & \cellcolor{blue!20}$s$ & $e$ & \cellcolor{yellow!30}$s^{-1}$ & \cellcolor{yellow!30}$d$ & \cellcolor{yellow!30}$f$ & \cellcolor{yellow!30}$ov^{-1}$ & \cellcolor{yellow!30}$m^{-1}$ & \cellcolor{yellow!30}$b^{-1}$ \\
\hline
\cellcolor{gray!8}$s^{-1}$ &\cellcolor{blue!20}$\beta_2$ & \cellcolor{blue!20}$\alpha_2$ & \cellcolor{blue!20}$\alpha_2$ & \cellcolor{blue!20}$d^{-1}$ & \cellcolor{blue!20}$d^{-1}$ & $\alpha_5$ & \cellcolor{yellow!30}$s^{-1}$ & \cellcolor{yellow!30}$s^{-1}$ & \cellcolor{yellow!30}$\alpha_2^{-1}$ & \cellcolor{yellow!30}$ov^{-1}$ & \cellcolor{yellow!30}$ov^{-1}$ & \cellcolor{yellow!30}$m^{-1}$ & \cellcolor{yellow!30}$b^{-1}$ \\
\hline
\cellcolor{gray!8}$d$ & \cellcolor{blue!20}$b$ & \cellcolor{blue!20}$b$ & \cellcolor{blue!20}$\beta_1$ & \cellcolor{blue!20}$\beta_1$ & $\delta$ & \cellcolor{yellow!30}$d$ & \cellcolor{yellow!30}$d$ & \cellcolor{yellow!30}$\beta_2^{-1}$ & \cellcolor{yellow!30}$d$ & \cellcolor{yellow!30}$d$ & \cellcolor{yellow!30}$\beta_2^{-1}$ & \cellcolor{yellow!30}$b^{-1}$ &\cellcolor{yellow!30} $b^{-1}$ \\
\hline
\cellcolor{gray!8}$f$ & \cellcolor{blue!20}$b$ & \cellcolor{blue!20}$m$ & \cellcolor{blue!20}$\alpha_1$ & $\alpha_4$ & \cellcolor{yellow!30}$\beta_1^{-1}$ & \cellcolor{yellow!30}$d$ & \cellcolor{yellow!30}$f$ & \cellcolor{yellow!30}$\alpha_3^{-1}$ & \cellcolor{yellow!30}$d$ & \cellcolor{yellow!30}$f$ & \cellcolor{yellow!30}$\alpha_3^{-1}$ & \cellcolor{yellow!30}$b^{-1}$ & \cellcolor{yellow!30}$b^{-1}$ \\
\hline
\cellcolor{gray!8}$ov^{-1}$ & \cellcolor{blue!20}$\beta_2$ & \cellcolor{blue!20}$\alpha_2$ & $\gamma$ & \cellcolor{yellow!30}$\alpha_1$ & \cellcolor{yellow!30}$\beta_1^{-1}$ & \cellcolor{yellow!30}$\alpha_2^{-1}$ & \cellcolor{yellow!30}$ov^{-1}$ & \cellcolor{yellow!30}$\alpha_3^{-1}$ & \cellcolor{yellow!30}$\alpha_2^{-1}$ & \cellcolor{yellow!30}$ov^{-1}$ & \cellcolor{yellow!30}$\alpha_3^{-1}$ & \cellcolor{yellow!30}$b^{-1}$ & \cellcolor{yellow!30} $b^{-1}$ \\
\hline
\cellcolor{gray!8}$m^{-1}$ & \cellcolor{blue!20}$\beta_2$ & $\alpha_5$ & \cellcolor{yellow!30}$\alpha_2^{-1}$ &\cellcolor{yellow!30} $m^{-1}$ & \cellcolor{yellow!30}$b^{-1}$ & \cellcolor{yellow!30}$\alpha_2^{-1}$ & \cellcolor{yellow!30}$m^{-1}$ & \cellcolor{yellow!30}$b^{-1}$ & \cellcolor{yellow!30}$\alpha_2^{-1}$ & \cellcolor{yellow!30}$m^{-1}$ & \cellcolor{yellow!30}$b^{-1}$ & \cellcolor{yellow!30}$b^{-1}$ & \cellcolor{yellow!30}$b^{-1}$ \\
\hline
\cellcolor{gray!8}$b^{-1}$ & $\delta$ & \cellcolor{yellow!30}$\beta_2^{-1}$ & \cellcolor{yellow!30}$\beta_2^{-1}$ & \cellcolor{yellow!30}$b^{-1}$ & \cellcolor{yellow!30}$b^{-1}$ & \cellcolor{yellow!30}$\beta_2^{-1}$ & \cellcolor{yellow!30}$b^{-1}$ & \cellcolor{yellow!30}$b^{-1}$ &\cellcolor{yellow!30} $\beta_2^{-1}$ & \cellcolor{yellow!30}$b^{-1}$ & \cellcolor{yellow!30}$b^{-1}$ & \cellcolor{yellow!30}$b^{-1}$ & \cellcolor{yellow!30}$b^{-1}$ \\
\hline
\end{tabular}
\end{center}
\end{table*}

\begin{figure}
\centering
\begin{tikzpicture}[line cap=round,line join=round,>=triangle 45,x=1.0cm,y=1.0cm, scale = 1]
\clip(-1,-1) rectangle (5,5);
\tikzstyle{every node}=[draw,shape=circle,radius = 0.06, fill=white];
\node (b) at (0,0) {}; 
\node (m) at (0,1) {}; 
\node (ov) at (0,2) {}; 
\node (fi) at (0,3) {}; 
\node (di) at (0,4) {}; 
\node (s) at (1,2) {}; 
\node (d) at (2,2) {}; 
\node (e) at (1,3) {}; 
\node (f) at (2,3) {}; 
\node (si) at (1,4) {}; 
\node (oi) at (2,4) {}; 
\node (mi) at (3,4) {}; 
\node (bi) at (4,4) {}; 

\draw {(b) -- (m) -- (ov) -- (fi) -- (di) -- (si) -- (oi)};
\draw {(ov) -- (s) -- (d) -- (f) -- (oi) -- (mi) -- (bi)};
\draw {(s) -- (e) -- (si)};
\draw {(fi) -- (e) -- (f)};
\tikzstyle{every node}=[draw=none,fill=none];
\node [ left] at (-0.2,0) {\small b};
\node [ left] at (-0.2,1) {\small m};
\node [ left] at (-0.2,2) {\small ov};
\node [ left] at (-0.2,3) {\small f$^{-1}$};
\node [ left] at (-0.2,4) {\small d$^{-1}$};
\node [below right] at (1.1,2) {\small s};
\node [below right] at (1.1,3) {\small e};
\node [above ] at (1,4.2) {\small s$^{-1}$};
\node [above ] at (2,4.2) {\small ov$^{-1}$};
\node [above ] at (3,4.2) {\small m$^{-1}$};
\node [above ] at (4,4.2) {\small b$^{-1}$};
\node [ right] at (2.1,2) {\small d};
\node [ right] at (2.1,3) {\small f};
\end{tikzpicture}
\caption{Lattice of time interval basic relations}
\label{fig:lattice}
\end{figure}
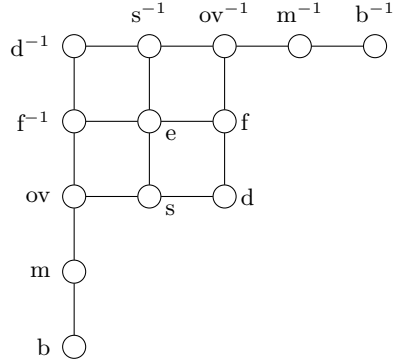
\subsection{Lattice Representation}
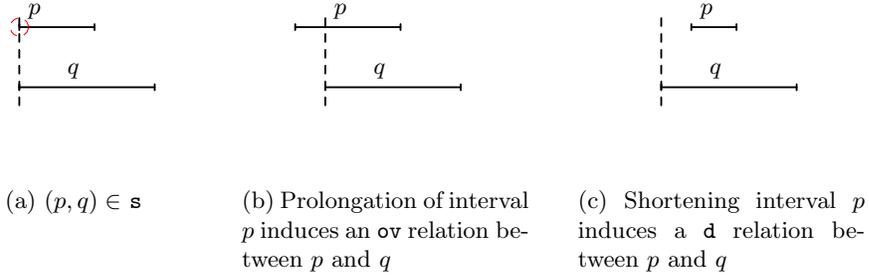
\begin{figure}
\centering
\subfloat[$(p,q)\in$ \texttt{s}]{
\begin{tikzpicture}[line cap=round,line join=round,>=triangle 45,x=0.8cm,y=0.8cm, scale = 0.5]
\clip(-0.4,-2) rectangle (7,4);

\draw [line width = 0.7pt] (1.5,3) -- (4,3);
\draw [line width = 0.7pt] (1.5,3.1) -- (1.5,2.9);
\draw [line width = 0.7pt] (4,3.1) -- (4,2.9);
\node [above] at (2,3) {$p$};

\draw [line width = 0.7pt] (1.5,1) -- (6,1);
\draw [line width = 0.7pt] (1.5,1.1) -- (1.5,0.9);
\draw [line width = 0.7pt] (6,1.1) -- (6,0.9);
\node [above] at (3.3,1) {$q$};

\draw [dashed, line width=0.7pt] (1.5,0.4) -- (1.5,3.4);
\draw[dashed, red] (1.5,3) circle [radius = 0.3];

\end{tikzpicture}
\label{subfig:cs}
}\qquad
\subfloat[Prolongation of interval $p$ induces an \texttt{ov} relation between $p$ and $q$]{
\begin{tikzpicture}[line cap=round,line join=round,>=triangle 45,x=0.8cm,y=0.8cm, scale = 0.5]
\clip(-1,-2) rectangle (8,4);
\draw [line width = 0.7pt] (0.5,3) -- (4,3);
\draw [line width = 0.7pt] (0.5,3.1) -- (0.5,2.9);
\draw [line width = 0.7pt] (4,3.1) -- (4,2.9);
\node [above] at (2,3) {$p$};

\draw [line width = 0.7pt] (1.5,1) -- (6,1);
\draw [line width = 0.7pt] (1.5,1.1) -- (1.5,0.9);
\draw [line width = 0.7pt] (6,1.1) -- (6,0.9);
\node [above] at (3.3,1) {$q$};

\draw [dashed, line width=0.7pt] (1.5,0.4) -- (1.5,3.4);

\end{tikzpicture}
\label{subfig:cov}
} \qquad
\subfloat[Shortening interval $p$ induces a \texttt{d} relation between $p$ and $q$]{
\begin{tikzpicture}[line cap=round,line join=round,>=triangle 45,x=0.8cm,y=0.8cm, scale = 0.5]
\clip(-1,-2) rectangle (8,4);
\draw [line width = 0.7pt] (2.5,3) -- (4,3);
\draw [line width = 0.7pt] (2.5,3.1) -- (2.5,2.9);
\draw [line width = 0.7pt] (4,3.1) -- (4,2.9);
\node [above] at (3,3) {$p$};

\draw [line width = 0.7pt] (1.5,1) -- (6,1);
\draw [line width = 0.7pt] (1.5,1.1) -- (1.5,0.9);
\draw [line width = 0.7pt] (6,1.1) -- (6,0.9);
\node [above] at (3.3,1) {$q$};
\draw [dashed, line width=0.7pt] (1.5,0.4) -- (1.5,3.4);
\end{tikzpicture}
\label{subfig:cd}
}
\caption{Possible scenarios of intervals $p$ and $q$ when translating the starting point of $p$ while keeping its ending point fixed}
\label{fig:csdov}
\end{figure}

A nice representation of the basic relations is the lattice representation~\cite{Lizgot:1992} depicted in Fig.~\ref{fig:lattice}. 
We say two relations form a \textit{conceptual neighborhood} \cite{Freksa:1992} if they are directly path-connected in the lattice in Fig.~\ref{fig:lattice}. From a topological point of view, it means that the two relations can be transformed into one another by either shortening or prolonging one interval.  Let $(p,q) \in $ \texttt{s}  as depicted in Fig.~\ref{subfig:cs}. To prolong interval $p$, we translate its starting point  $p$ to the left while keeping the ending point fixed  as shown in Fig.~\ref{subfig:cov}. We obtain a new configuration where $(p,q) \in $ \texttt{ov}. Now, we shorten $p$ by translating the starting point to the right as shown in Fig.~\ref{subfig:cd}. Then, $(p,q) \in $ \texttt{d}. 

The three relations \texttt{s}, \texttt{ov} and \texttt{d} are deduced from the configurations of the starting points of $p$ and $q$. If we apply (M2) to literals that constraint the starting points, which means literals of the form $x||p$ and $y||q$, then we obtain the three exclusive cases $x||q\; \oplus\; (\exists t.~x||t \land t||q)\; \oplus \;(\exists t.~y||t \land t||p)$. 
These three cases correspond to Fig.~\ref{subfig:cs}, \ref{subfig:cd} and  \ref{subfig:cov}, respectively.

 We observe that the relations $\alpha$, $\beta$, $\gamma$ and $\delta$ are union of relations that are path-connected in the lattice. The ordering of their relations given by the lattice turns out to be extremely useful for proving them. 

\section{Proof of Validity of the Composition Table}
\label{sect:proof}
\subsection{Automation}
What we need to prove is the entries in blue and white cells in Table \ref{table1}. The rest is deduced straightforwardly. A complete proof of the composition table has been done in Isabelle/HOL and our theory files are available at \cite{Fadoua:2016}.

For now, we focus on the blue cells that contain single relation. The compositions $r_1 \circ e \subseteq r_1$, $e \circ r_1 \subseteq r_1$, $r_1 \circ r_2 \subseteq \texttt{b}$ and $r_1 \circ r_2 \subseteq \texttt{m}$ are proved by applying \texttt{blast} with axioms (M1), (M4) and (M5).  In our experiments, the \texttt{blast} tactic as well as ATP systems and SMT solvers invoked by Isabelle/HOL were not able to prove the compositions $r_1 \circ r_2 \subseteq \texttt{ov}$, $r_1 \circ r_2 \subseteq \texttt{f}^{-1}$, $r_1 \circ r_2 \subseteq \texttt{d}^{-1}$ and $r_1 \circ r_2 \subseteq \texttt{s}$. 
We then prove these compositions in forward proof style with Isar. The literals of the result relation are inferred from the literals in $r_1$ and $r_2$. To that end, axioms (M1) and (M5) are all what we need and they are called more than once. This means that the number of candidate literals  increases in vain if we do not control the way we apply (M1) and (M5). 
A forward proof is performed by the careful choice of the ``right" literals to infer new ones with (M1) and (M5) until we collect all the literals that satisfy the result relation. 


Regarding the proof of compositions $\alpha, \beta, \gamma$ and $\delta$, ATP systems  and SMT solvers gave up and were not able to automate the proof. We thus write a structured proof in Isar for those relations, which is covered in more details in Sect.~\ref{sect:alpha} $\sim$ \ref{sect:gammadelta}. 
\subsection{Proof Goals}
Proving compositions $\alpha, \beta, \gamma$ and $\delta$ means showing lemmas of the form
$$r_1 \circ r_2 \subseteq \theta_1 \cup \ldots \cup \theta_n\text{, where $\theta_{1\leq n \leq 13}$ are basic relations}$$ 

which can be simplified to
$$\forall p~q. ~ (p,q)\in r_1 \circ r_2 \longrightarrow (p,q) \in \theta_1 \lor\ldots \lor (p,q) \in \theta_n\text{, where $1\leq n \leq 13$}$$

The above goal is weak. Due to the disjunctive conclusion, a proof of one $(p,q) \in \theta_i$ is enough. But, we rather aim at showing situations of $p$ and $q$ where  the expressions $(p,q) \in \theta_{i}$ hold for all $1\leq i \leq n$. We aim at finding witnesses $z$ that satisfy the following  subgoals. 
$$\exists z.~(p,z)\in r_1 \land (z,q)\in r_2 \land (p,q) \in \theta_1$$
\vskip -2.2em
$$\ldots$$
\vskip -2em
$$\exists z.~(p,z)\in r_1 \land (z,q)\in r_2 \land (p,q) \in \theta_i$$
\vskip -2em
$$\ldots$$
$$\exists z.~(p,z)\in r_1 \land (z,q)\in r_2 \land (p,q) \in \theta_n$$
\vskip -2em

\subsection{Proof  of  $\alpha$ Composition}
\label{sect:alpha}


Each of the relation $\alpha_i$ is union of three basic relations that are path-connected in  Fig.~\ref{fig:lattice}. One of the basic relation form a conceptual neighbourhood with the two others. In $\alpha_1 = (\texttt{ov}, \texttt{s}, \texttt{d})$, conceptual neighborhood is given by relations \texttt{s} and \texttt{d} in one hand, and \texttt{s} and \texttt{ov} on the other hand. Recall that axiom (M2) is linked to the conceptual neighborhood in the sense it allows deducing three exclusive configurations that correspond to  three path-connected relations.  Proving an $\alpha$ composition boils down to applying axiom (M2) on two suitable literals.  
%
The three relations \texttt{s}, \texttt{ov} and \texttt{d} are deduced from translating the starting point of interval $p$ as depicted in Fig.~\ref{fig:csdov}. Accordingly, the constraints on the starting points of $p$ and $q$ change. We therefore apply (M2) to literals of the form $x||p$ and $y||q$. 

\begin{figure}
\centering
\subfloat[$(p,q)\in$ \texttt{e}]{
\begin{tikzpicture}[line cap=round,line join=round,>=triangle 45,x=0.8cm,y=0.8cm, scale = 0.5]
\clip(-0.4,-2) rectangle (7,4);

\draw [line width = 0.7pt] (2,3) -- (4,3);
\draw [line width = 0.7pt] (2,3.1) -- (2,2.9);
\draw [line width = 0.7pt] (4,3.1) -- (4,2.9);
\node [above] at (3,3) {$p$};

\draw [line width = 0.7pt] (2,1) -- (4,1);
\draw [line width = 0.7pt] (2,1.1) -- (2,0.9);
\draw [line width = 0.7pt] (4,1.1) -- (4,0.9);
\node [above] at (3,1) {$q$};

\draw [dashed, line width=0.7pt] (4,0.4) -- (4,3.4);
\draw [dashed, line width=0.7pt] (2,0.4) -- (2,3.4);
\draw[dashed, red] (2,3) circle [radius = 0.3];

\end{tikzpicture}
\label{subfig:ce}
}\qquad
\subfloat[Prolongation of starting point of $p$ induces $(p,q)\in$ \texttt{f}$^{-1}$]{
\begin{tikzpicture}[line cap=round,line join=round,>=triangle 45,x=0.8cm,y=0.8cm, scale = 0.5]
\clip(-1,-2) rectangle (8,4);
\draw [line width = 0.7pt] (0.5,3) -- (4,3);
\draw [line width = 0.7pt] (0.5,3.1) -- (0.5,2.9);
\draw [line width = 0.7pt] (4,3.1) -- (4,2.9);
\node [above] at (3,3) {$p$};

\draw [line width = 0.7pt] (2,1) -- (4,1);
\draw [line width = 0.7pt] (2,1.1) -- (2,0.9);
\draw [line width = 0.7pt] (4,1.1) -- (4,0.9);
\node [above] at (3,1) {$q$};

\draw [dashed, line width=0.7pt] (4,0.4) -- (4,3.4);
\draw [dashed, line width=0.7pt] (2,0.4) -- (2,3.4);

\end{tikzpicture}
\label{subfig:cefi}
} \qquad
\subfloat[Shortening the starting point of $p$ induces $(p,q) \in $ \texttt{f}]{
\begin{tikzpicture}[line cap=round,line join=round,>=triangle 45,x=0.8cm,y=0.8cm, scale = 0.5]
\clip(-1,-2) rectangle (8,4);
\draw [line width = 0.7pt] (2.8,3) -- (4,3);
\draw [line width = 0.7pt] (2.8,3.1) -- (2.8,2.9);
\draw [line width = 0.7pt] (4,3.1) -- (4,2.9);
\node [above] at (3.5,3) {$p$};

\draw [line width = 0.7pt] (2,1) -- (4,1);
\draw [line width = 0.7pt] (2,1.1) -- (2,0.9);
\draw [line width = 0.7pt] (4,1.1) -- (4,0.9);
\node [above] at (3,1) {$q$};

\draw [dashed, line width=0.7pt] (4,0.4) -- (4,3.4);
\draw [dashed, line width=0.7pt] (2,0.4) -- (2,3.4);

\end{tikzpicture}
\label{subfig:cef}
}
\caption{The proof of $(p,q) \in \alpha_4$ requires applying (M2) on two literals of the form $x||p$ and $y||q$}
\label{fig:cefif}
\end{figure}
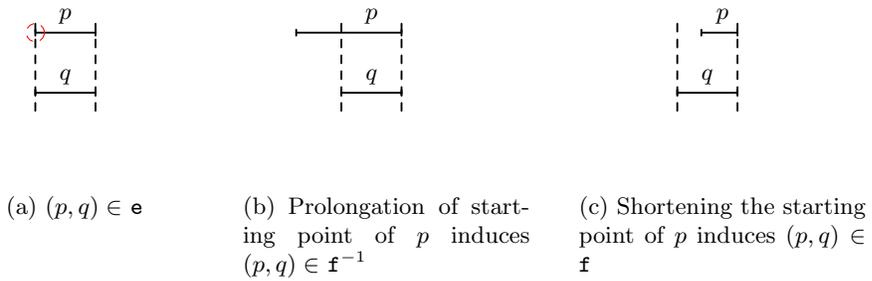

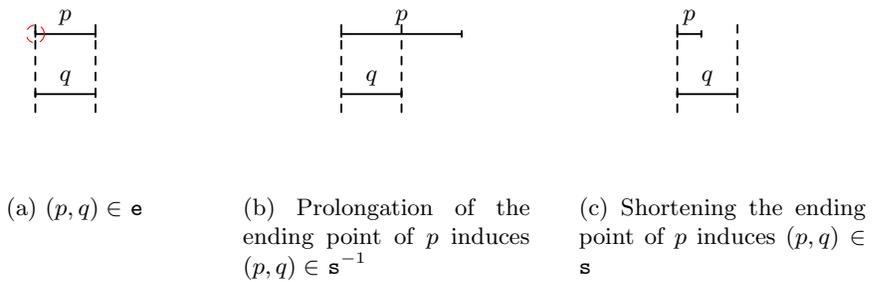
\begin{figure}
\centering
\subfloat[$(p,q)\in$ \texttt{e}]{
\begin{tikzpicture}[line cap=round,line join=round,>=triangle 45,x=0.8cm,y=0.8cm, scale = 0.5]
\clip(-0.4,-2) rectangle (7,4);

\draw [line width = 0.7pt] (2,3) -- (4,3);
\draw [line width = 0.7pt] (2,3.1) -- (2,2.9);
\draw [line width = 0.7pt] (4,3.1) -- (4,2.9);
\node [above] at (3,3) {$p$};

\draw [line width = 0.7pt] (2,1) -- (4,1);
\draw [line width = 0.7pt] (2,1.1) -- (2,0.9);
\draw [line width = 0.7pt] (4,1.1) -- (4,0.9);
\node [above] at (3,1) {$q$};

\draw [dashed, line width=0.7pt] (4,0.4) -- (4,3.4);
\draw [dashed, line width=0.7pt] (2,0.4) -- (2,3.4);
\draw[dashed, red] (2,3) circle [radius = 0.3];

\end{tikzpicture}
\label{subfig:ce2}
}\qquad
\subfloat[Prolongation of the ending point of $p$ induces $(p,q)\in$ \texttt{s}$^{-1}$]{
\begin{tikzpicture}[line cap=round,line join=round,>=triangle 45,x=0.8cm,y=0.8cm, scale = 0.5]
\clip(-1,-2) rectangle (8,4);
\draw [line width = 0.7pt] (2,3) -- (6,3);
\draw [line width = 0.7pt] (2,3.1) -- (2,2.9);
\draw [line width = 0.7pt] (6,3.1) -- (6,2.9);
\node [above] at (4,3) {$p$};
\draw [line width = 0.7pt] (2,1) -- (4,1);
\draw [line width = 0.7pt] (2,1.1) -- (2,0.9);
\draw [line width = 0.7pt] (4,1.1) -- (4,0.9);
\node [above] at (3,1) {$q$};

\draw [dashed, line width=0.7pt] (4,0.4) -- (4,3.4);
\draw [dashed, line width=0.7pt] (2,0.4) -- (2,3.4);

\end{tikzpicture}
\label{subfig:cesi}
} \qquad
\subfloat[Shortening the ending point of $p$ induces $(p,q) \in $ \texttt{s}]{
\begin{tikzpicture}[line cap=round,line join=round,>=triangle 45,x=0.8cm,y=0.8cm, scale = 0.5]
\clip(-1,-2) rectangle (8,4);
\draw [line width = 0.7pt] (2,3) -- (2.8,3);
\draw [line width = 0.7pt] (2,3.1) -- (2,2.9);
\draw [line width = 0.7pt] (2.8,3.1) -- (2.8,2.9);
\node [above] at (2.4,3) {$p$};

\draw [line width = 0.7pt] (2,1) -- (4,1);
\draw [line width = 0.7pt] (2,1.1) -- (2,0.9);
\draw [line width = 0.7pt] (4,1.1) -- (4,0.9);
\node [above] at (3,1) {$q$};

\draw [dashed, line width=0.7pt] (4,0.4) -- (4,3.4);
\draw [dashed, line width=0.7pt] (2,0.4) -- (2,3.4);

\end{tikzpicture}
\label{subfig:ces}
}
\caption{The proof of $(p,q) \in \alpha_5$ requires applying (M2) on two literals of the form $p||x$ and $q||y$}
\label{fig:cesis}
\end{figure}

To prove $\alpha_2 \sim \alpha_5$ compositions, we also single out suitable literals to apply (M2) and deduce the three configurations that correspond to  three relations. 
Figures \ref{fig:csfidiov}, \ref{fig:cbmov}, \ref{fig:cefif} and \ref{fig:cesis} show the application of (M2) to prove $\alpha_2$, $\alpha_3$, $\alpha_4$ and $\alpha_5$, respectively.

The following is a general template for proving the compositions $r_1 \circ r_2 \subseteq \alpha_i$, where $1\leq i \leq 5$ and $\alpha_i = ({\theta_i}_1, {\theta_i}_2, {\theta_i}_3)$. 

\begin{flushleft}
\textbf{Proof template}  [Proof of $r_1 \circ  r_2 \subseteq \alpha_i$]
\end{flushleft}
\begin{itemize}
\item[1.] Let $p$, $z$ and $q$ be three intervals where $(p,z) \in r_1$, $(z,q) \in r_2$ and $(p,q) \in r_1 \circ r_2$ 
\item[2.] Obtain literals  $L_{r_1}(p,z)$ and $L_{r_2}(z,q)$
\item[3.] From $L_{r_1}(p,z)$ and $L_{r_2}(z,q)$, single out two suitable literals depending on $\alpha_i$
\begin{itemize}
\item[(a)] Choose two literals of the form $x||p$ and $y||q$ if $i = 1$ or $i=4$
\item[(b)] Choose two literals of the form $p||x$ and $q||y$ if $i= 2$ or $i=5$
\item[(c)] Choose two literals of the form $p||x$ and $y||q$ if $i = 3$
\end{itemize}
\item[4.] If there is no literals to choose in the previous step, then apply (M3) to generate them

\item[5.] Apply (M2) on $x||p$ and $y||q$ 
\begin{itemize}
\item[] (Case $x||q$) Deduce all literals in $L_{{\theta_i}_1}(p,q)$ using (M1), (M4) and (M5) when necessary
\item[] (Case $\exists t.\;y||t\land t||p$)
Deduce all literals in $L_{{\theta_i}_2}(p,q)$ using (M1), (M4) and (M5) when necessary  
\item[] (Case $\exists t.\;x||t\land t||q$)
Deduce all literals in $L_{{\theta_i}_3}(p,q)$ using (M1), (M4) and (M5) when necessary 
\end{itemize}
\end{itemize}

As example, we explain the proof of the following lemma.
\begin{isabellebody}
\isanewline
\isacommand{lemma}\isamarkupfalse%
\ \ {\isachardoublequoteopen}m\ O\ d\ {\isasymsubseteq}\ s\ {\isasymunion}\ ov\ {\isasymunion}\ d{\isachardoublequoteclose}
\end{isabellebody}
\begin{proof}
\hfill
\begin{itemize}
\item[1.] Let $p$, $z$ and $q$ be intervals where $(p,z) \in m$, $(z,q) \in d$ and $(p,q) \in$ m $\circ$ d.
\item[2.] We have $L_{\texttt{m}}(p,z) = \{p||z\}$ and $L_{\texttt{d}}(z,q)=\{k||q, k||l, l||z, z||u, u||v, q||v\}$.
\item[3.] We single out $k||q$.
\item[4.] We apply (M3) to obtain an interval $c$ where $c||p$.
\item[5.] We apply (M2) on $c||p$ and $k||q$.
\begin{itemize}
\item[] (Case $c||q$) 
\begin{itemize}
\item We apply (M5) to add $z$ and $u$ and obtain new time interval $zu$, i.e. $\llbracket p||z; \;z||u; \; u||v \rrbracket \Longrightarrow \exists zu.\;p||zu \land zu||v$.
\item All literals in $L_\texttt{s}(p,q) = \{c||p, x||q,p||zu, zu||v, q||v\}$ are obtained. We deduce $(p,q) \in \texttt{s}$.
\end{itemize}
\item[] (Case  $\exists t.\;k||t\land t||p$)
\begin{itemize}
\item We apply (M5) to add $z$ and $u$ and obtain new time interval $zu$, i.e. $\llbracket p||z; \; z||u; \; u||v \rrbracket \Longrightarrow \exists zu.\;p||zu \land zu||v$.
\item All literals in $L_\texttt{d}(p,q) = \{k||t, t||p, p||zu, zu||v, k||q, q||v\}$ are obtained. We deduce $(p,q) \in \texttt{d}$. 
\end{itemize}
\item[] (Case $\exists t.\;c||t\land t||q$)
\begin{itemize}
\item We apply (M5) to add $z$ and $u$ and obtain new time interval $zu$, i.e. $\llbracket p||z ; \; z||u ; \; u||v \rrbracket \Longrightarrow \exists zu.\;p||zu \land zu||v$.
\item We apply (M1) to deduce $t||l$, i.e.~$\llbracket k||q ; \; k||l; \; t||q \rrbracket \Longrightarrow t||l$.
\item We apply (M1) to deduce $l||zu$, i.e.~$\llbracket l||z ; \; p||z ; \; p||zu \rrbracket \Longrightarrow l||zu$.
\item All literals in $L_{\texttt{ov}}(p,q) = \{c||p, p||zu, zu||v, c||t, t||q, q||v, t||l, l||zu\}$ are obtained. We deduce $(p,q) \in \texttt{ov}$.
\end{itemize}
\end{itemize}
\end{itemize}
\end{proof}
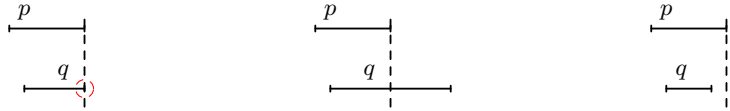
\begin{figure}
\centering
\subfloat[$(p,q)\in$ \texttt{f}$^{-1}$]{
\begin{tikzpicture}[line cap=round,line join=round,>=triangle 45,x=0.8cm,y=0.8cm, scale = 0.5]
\clip(-0.4,-2) rectangle (7,4);

\draw [line width = 0.7pt] (1.5,3) -- (4,3);
\draw [line width = 0.7pt] (1.5,3.1) -- (1.5,2.9);
\draw [line width = 0.7pt] (4,3.1) -- (4,2.9);
\node [above] at (2,3) {$p$};

\draw [line width = 0.7pt] (2,1) -- (4,1);
\draw [line width = 0.7pt] (2,1.1) -- (2,0.9);
\draw [line width = 0.7pt] (4,1.1) -- (4,0.9);
\node [above] at (3.3,1) {$q$};

\draw [dashed, line width=0.7pt] (4,0.4) -- (4,3.4);
\draw[dashed, red] (4,1) circle [radius = 0.3];

\end{tikzpicture}
\label{subfig:cfi}
}\qquad
\subfloat[Prolongation of interval $q$ induces an \texttt{ov} relation between $p$ and $q$]{
\begin{tikzpicture}[line cap=round,line join=round,>=triangle 45,x=0.8cm,y=0.8cm, scale = 0.5]
\clip(-1,-2) rectangle (8,4);
\draw [line width = 0.7pt] (1.5,3) -- (4,3);
\draw [line width = 0.7pt] (1.5,3.1) -- (1.5,2.9);
\draw [line width = 0.7pt] (4,3.1) -- (4,2.9);
\node [above] at (2,3) {$p$};
\draw [line width = 0.7pt] (2,1) -- (6,1);
\draw [line width = 0.7pt] (2,1.1) -- (2,0.9);
\draw [line width = 0.7pt] (6,1.1) -- (6,0.9);
\node [above] at (3.3,1) {$q$};

\draw [dashed, line width=0.7pt] (4,0.4) -- (4,3.4);

\end{tikzpicture}
\label{subfig:cfov}
} \qquad
\subfloat[Shortening interval $p$ induces a \texttt{d}$^{-1}$ relation between $p$ and $q$]{
\begin{tikzpicture}[line cap=round,line join=round,>=triangle 45,x=0.8cm,y=0.8cm, scale = 0.5]
\clip(-1,-2) rectangle (8,4);
\draw [line width = 0.7pt] (1.5,3) -- (4,3);
\draw [line width = 0.7pt] (1.5,3.1) -- (1.5,2.9);
\draw [line width = 0.7pt] (4,3.1) -- (4,2.9);
\node [above] at (2,3) {$p$};

\draw [line width = 0.7pt] (2,1) -- (3.5,1);
\draw [line width = 0.7pt] (2,1.1) -- (2,0.9);
\draw [line width = 0.7pt] (3.5,1.1) -- (3.5,0.9);
\node [above] at (2.5,1) {$q$};
\draw [dashed, line width=0.7pt] (4,0.4) -- (4,3.4);
\end{tikzpicture}
\label{subfig:cfd}
}
\caption{The proof of $(p,q) \in \alpha_2$ requires applying (M2) on two literals of the form $p||x$ and $q||y$}
\label{fig:csfidiov}
\end{figure}

\begin{figure}
\centering
\subfloat[$(p,q)\in$ \texttt{m}]{
\begin{tikzpicture}[line cap=round,line join=round,>=triangle 45,x=0.8cm,y=0.8cm, scale = 0.5]
\clip(-0.4,-2) rectangle (7,4);

\draw [line width = 0.7pt] (1.5,3) -- (4,3);
\draw [line width = 0.7pt] (1.5,3.1) -- (1.5,2.9);
\draw [line width = 0.7pt] (4,3.1) -- (4,2.9);
\node [above] at (2,3) {$p$};

\draw [line width = 0.7pt] (4,1) -- (6,1);
\draw [line width = 0.7pt] (4,1.1) -- (4,0.9);
\draw [line width = 0.7pt] (6,1.1) -- (6,0.9);
\node [above] at (5.3,1) {$q$};

\draw [dashed, line width=0.7pt] (4,0.4) -- (4,3.4);
\draw[dashed, red] (4,1) circle [radius = 0.3];

\end{tikzpicture}
\label{subfig:cm}
}\qquad
\subfloat[Prolongation of starting point of $q$ induces $(p,q)\in$ \texttt{ov}]{
\begin{tikzpicture}[line cap=round,line join=round,>=triangle 45,x=0.8cm,y=0.8cm, scale = 0.5]
\clip(-1,-2) rectangle (8,4);
\draw [line width = 0.7pt] (1.5,3) -- (4,3);
\draw [line width = 0.7pt] (1.5,3.1) -- (1.5,2.9);
\draw [line width = 0.7pt] (4,3.1) -- (4,2.9);
\node [above] at (2,3) {$p$};
\draw [line width = 0.7pt] (2,1) -- (6,1);
\draw [line width = 0.7pt] (2,1.1) -- (2,0.9);
\draw [line width = 0.7pt] (6,1.1) -- (6,0.9);
\node [above] at (3.3,1) {$q$};

\draw [dashed, line width=0.7pt] (4,0.4) -- (4,3.4);

\end{tikzpicture}
\label{subfig:cmov}
} \qquad
\subfloat[Shortening the starting point of $q$ induces $(p,q) \in $ \texttt{b}]{
\begin{tikzpicture}[line cap=round,line join=round,>=triangle 45,x=0.8cm,y=0.8cm, scale = 0.5]
\clip(-1,-2) rectangle (8,4);
\draw [line width = 0.7pt] (1.5,3) -- (4,3);
\draw [line width = 0.7pt] (1.5,3.1) -- (1.5,2.9);
\draw [line width = 0.7pt] (4,3.1) -- (4,2.9);
\node [above] at (2,3) {$p$};

\draw [line width = 0.7pt] (4.5,1) -- (6,1);
\draw [line width = 0.7pt] (4.5,1.1) -- (4.5,0.9);
\draw [line width = 0.7pt] (6,1.1) -- (6,0.9);
\node [above] at (5,1) {$q$};
\draw [dashed, line width=0.7pt] (4,0.4) -- (4,3.4);
\end{tikzpicture}
\label{subfig:cmb}
}
\caption{The proof of $(p,q) \in \alpha_3$ requires applying (M2) on two literals of the form $p||x$ and $y||q$}
\label{fig:cbmov}
\end{figure}
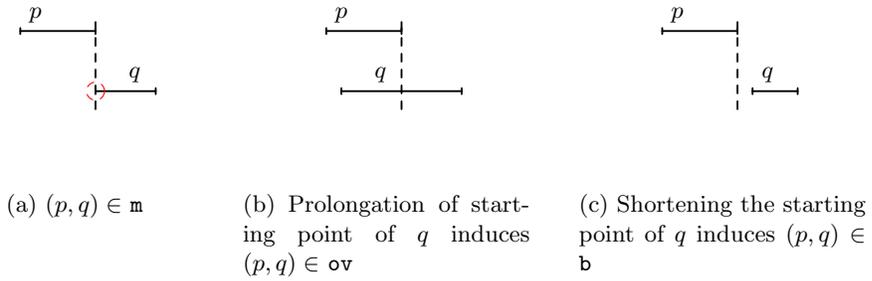

\subsection{Proof of $\beta$ Compositions}
\label{sect:beta}

We have relations $\beta_1 = (\texttt{b}, \texttt{m}, \texttt{ov}, \texttt{s}, \texttt{d})$. We notice that $\beta_1 = \alpha_1 + \alpha_3$  and relation \texttt{ov} appears in $\alpha_1$ and $\alpha_3$.  This observation means that first we proceed to prove $\alpha_1$, then from the case that corresponds to \texttt{ov} configuration in the proof of $\alpha_1$, we   apply (M2) to deduce $\alpha_3$. Similarly for  $\beta_2 = (\texttt{b}, \texttt{m}, \texttt{ov}, \texttt{f}^{-1}, \texttt{d}^{-1})$ = $\alpha_2 + \alpha_3$. 

The proof template for the $\alpha$ composition is thus extended to perform the proof of $\beta$ composition. We only show  the proof template of $r_1 \circ r_2 \subseteq \beta_1$. The proof template of $r_1 \circ r_2 \subseteq \beta_2$ is written in the same fashion.

\begin{flushleft}
\textbf{Proof template}  [Proof of $r_1 \circ  r_2 \subseteq \beta_1$]
\end{flushleft}
\begin{enumerate}
\item Let $p$, $z$ and $q$ be three intervals where $(p,z) \in r_1$, $(z,q) \in r_2$ and $(p,q) \in r_1 \circ r_2$ 
\item Obtain literals  $L_{r_1}(p,z)$ and $L_{r_2}(z,q)$
\item From $L_{r_1}(p,z)$ and $L_{r_2}(z,q)$, single out two suitable literals of the form $x||p$ and $y||q$
\item If there is no literals to choose in the previous step, then apply (M3) to generate them

\item Apply (M2) on $x||p$ and $y||q$ 
\begin{itemize}
\item[] (Case $x||q$) Deduce all literals in $L_\texttt{s}(p,q)$using (M1), (M4) and (M5) when necessary. 
\item[] (Case $\exists t.\;y||t\land t||p$)
Deduce all literals in $L_\texttt{d}(p,q)$ using (M1), (M4) and (M5) when necessary.  
\item[] (Case $\exists t.\;x||t\land t||q$) 
\begin{itemize}
\item[5.1] Single out two suitable literals of the form $p||x'$ and $y'||q$
\item[5.2] If there is no literals to choose, then apply (M3) to generate them.
\item[5.3] Apply (M2) on $p||x'$ and $y'||q$
\begin{itemize}
\item[] (Case $p||q$) $(p,q) \in $ \texttt{m} is immediate. 
\item[] (Case $\exists t'.\;p||t'\land t'||q$) $(p,q) \in $ \texttt{b} is immediate.
\item[] (Case $\exists t'.\;y'||t'\land t'||x'$) Deduce all literals in $L_\texttt{ov}(p,q)$ using (M1), (M4) and (M5) when necessary.  
\end{itemize}
\end{itemize}
\end{itemize}
\end{enumerate}

\subsection{Proof of $\gamma$ and $\delta$ Compositions}
\label{sect:gammadelta}

Note that $\gamma = (\texttt{ov}, \texttt{s}, \texttt{d}, \texttt{f}, \texttt{e}, \texttt{f}^{-1}, \texttt{d}^{-1}, \texttt{s}^{-1}, \texttt{ov}^{-1}) = \alpha_1 +  \alpha_4  + {\alpha_1}^{-1}$. We apply (M2) on $p||x$ and $p||y$ to generate three cases. For each case, we split again with (M2) applied to $x'||p$ and $y'||q$ to generate $3\times3$  cases that leads to  the nine relations in $\gamma$. 

We have $\gamma = (\texttt{b}, \texttt{m}, \texttt{ov}, \texttt{s}, \texttt{d}, \texttt{f}, \texttt{e}, \texttt{f}^{-1}, \texttt{d}^{-1}, \texttt{s}^{-1}, \texttt{ov}^{-1}, \texttt{m}^{-1}, \texttt{b}^{-1}) = \alpha_3 +  \alpha_1 + \alpha_4  + {\alpha_1}^{-1} + \alpha_3^{-1}$. We follow the above strategy for $\gamma$ to deduce $\alpha_1$, $\alpha_1^{-1}$ and $\alpha_4$, except in the cases that lead to \texttt{ov} and \texttt{ov}$^{-1}$, we split again to deduce $\alpha_3$ and $\alpha_3^{-1}$. 

\section{Conclusion}
\label{sect:conc}
We proved the validity of composition table of Allen's calculus.\footnote{The proofs with Isabelle/HOL are available at {https://www.isa-afp.org/entries/Allen\_Calculus.html}}. We explained our strategy based on lattice structure of time interval relations. The axiom (M2) is central to browsing the lattice and consequently deducing the relations one by one. The proof of each entry of the composition table assumes what the composition will be (computed first  in \cite{Allen:1983}). The next challenge would be not to assume this. This is useful for constructing the compositions when designing extension of Allen's calculus.  
Another possible direction of this formalization would be to convert proof templates into a decision procedure for Allen's interval calculus. An interesting element would be the decision heuristic by which literals are chosen to apply the axioms.


\label{sect:bib}
\bibliographystyle{plain}
\bibliography{QSR}

\begin{thebibliography}{10}

\bibitem{Buchberger:1985}
B.~Buchberger.
\newblock {Groebner-Bases: An Algorithmic Method in Polynomial Ideal Theory}.
\newblock In {\em {Multidimensional Systems Theory - Progress, Directions and
  Open Problems in Multidimensional Systems}}, chapter~6, pages 184--232.
  Copyright: Reidel Publishing Company, Dordrecht - Boston - Lancaster, The
  Netherlands, 1985.

\bibitem{Collins:1996}
G.~Collins.
\newblock Quantifier elimination by cylindrical algebraic decomposition.
\newblock In H.~Brakhage, editor, {\em Automata theory and formal languages},
  volume~33 of {\em Lecture Notes in Computer Science}, pages 134--183, 1996.

\bibitem{Dylla:2013}
F.~Dylla, T.~Mossakowski, T.~Schneider, and D.~Wolter.
\newblock {Algebraic Properties of Qualitative Spatio-Temporal Calculi}.
\newblock In {\em {Proceedings of the 11th Conference on Spatial Information
  Theory (COSIT)}}, volume 8116 of {\em LNCS}, pages 516--536. Springer, 2013.

\bibitem{Frank:1991}
A.~U. Frank.
\newblock {Qualitative Spatial Reasoning about Cardinal Directions}.
\newblock In {\em {Proceedings of the International Symposium on
  Computer-Assisted Cartography}}, pages 148--167. ACSM-ASPRS, 1991.

\bibitem{Freksa:1992}
C.~Freksa.
\newblock Temporal reasoning based on semi-intervals.
\newblock {\em Artificial Intelligence}, 54(1-2):199 --227, 1992.

\bibitem{Fadoua:2016}
F.~Ghourabi.
\newblock {Isabelle/HOL Theories of Time Interval Relations}.
\newblock \url{http://www.i-eos.org/fadoua/IAtheories/}, June 2016.

\bibitem{Fadoua:2015}
F.~Ghourabi and K.~Takahashi.
\newblock {Formalizing the Qualitative Superposition of Rectangles in Proof
  Assistant Isabelle/HOL}.
\newblock In {\em Proceedings of the International Conference on Agents and
  Artificial Intelligence}, pages 530--539, 2015.

\bibitem{Allen:1985}
{J. F. Allen and P. J. Hayes}.
\newblock {A Common-sense Theory of Time}.
\newblock In {\em {Proceedings of the 9th International Joint Conference on
  Artificial Intelligence (IJCAI'85)}}, pages 528--531, 1985.

\bibitem{Allen:1983}
{J.F. Allen}.
\newblock {Maintaining Knowledge about Temporal Intervals}.
\newblock In {\em {Commun. ACM}}, volume~26, pages 832--843, 1983.

\bibitem{Ladkin:1992}
P.~B. Ladkin and A.~Reinefeld.
\newblock Effective solution of qualitative interval constraint problems.
\newblock {\em Artificial Intelligence}, 57(1):105 -- 124, 1992.

\bibitem{Lizgot:1992}
G.~Ligozat.
\newblock {\em {Qualitative Spatial and Temporal Reasoning}}.
\newblock John Wiley \& Sons, Inc., 2013.

\bibitem{Randell:1992a}
D.~A. Randell, Z.~Cui, and A.~G. Cohn.
\newblock {A Spatial Logic based on Regions and Connection}.
\newblock In {\em {Proceedings of the 3rd International Conference on Knowledge
  Representation and Reasoning}}, 1992.

\bibitem{Wallgrun:2007}
J.~O. Wallgr{\"u}n, L.~Frommberger, D.~Wolter, F.~Dylla, and C.~Freksa.
\newblock Qualitative spatial representation and reasoning in the
  sparq-toolbox.
\newblock In {\em Spatial Cognition V Reasoning, Action, Interaction:
  International Conference Spatial Cognition, Revised Selected Papers}, pages
  39--58, Berlin, Heidelberg, 2007. Springer Berlin Heidelberg.

\bibitem{Wolter:2012}
D.~Wolter.
\newblock {Analyzing Qualitative Spatio-Temporal Calculi using Algebraic
  Geometry}.
\newblock {\em {Spatial Cognition \& Computation}}, 12(1):23--52, 2012.

\end{thebibliography}
\end{document}